\documentclass[11pt]{article}
\input epsf.tex
\newdimen\SaveWidth \SaveWidth=\textwidth
\newdimen\SaveHeight \SaveHeight=\textheight
\advance\SaveWidth by -\textwidth
\advance\SaveHeight by -\textheight
\divide\SaveWidth by 2
\divide\SaveHeight by 2
\advance\hoffset by \SaveWidth
\advance\voffset by \SaveHeight

\def\sgn{\mathop{\rm sgn}}
\def\etmiss{{\slashchar{E}_T}}

\def\fbi{{\rm fb}^{-1}}
\def\Meff{M_{\rm eff}}

\def\ra{\rightarrow}
\def\GeV{{\rm GeV}}

\def\mhalf{m_{1/2}}

\def\lsp{{\tilde\chi_1^0}}
\def\tchi{{\tilde\chi}}
\def\tg{{\tilde g}}
\def\tq{{\tilde q}}
\def\tell{{\tilde\ell}}

\def\tG{{\tilde G}}

\def\rmax{{\rm max}}

\let\badcite=\cite
\def\cite{~\badcite}
\catcode`@=11
\def\ifundefined#1{\expandafter\ifx\csname#1\endcsname\relax}
\def\citenum#1{\ifundefined{b@#1}{\bf#1}%
   \immediate\write16{citenum: Undefined argument #1}%
   \else\csname b@#1\endcsname\fi}
\catcode`@=12

\def\slashchar#1{\setbox0=\hbox{$#1$}           % set a box for #1
   \dimen0=\wd0                                 % and get its size
   \setbox1=\hbox{/} \dimen1=\wd1               % get size of /
   \ifdim\dimen0>\dimen1                        % #1 is bigger
      \rlap{\hbox to \dimen0{\hfil/\hfil}}      % so center / in box
      #1                                        % and print #1
   \else                                        % / is bigger
      \rlap{\hbox to \dimen1{\hfil$#1$\hfil}}   % so center #1
      /                                         % and print /
   \fi}                                         %

\def\dofig#1#2{%
   \vskip-12pt%
   \centerline{\epsfxsize=#1\epsfbox{#2}}%
   \vskip-12pt}
\def\dofigs#1#2#3{%
   \vskip-12pt%
    \centerline{\epsfxsize=#1\epsfbox{#2}\hfil\epsfxsize=#1\epsfbox{#3}}%
   \vskip-12pt}

\begin{document}
\rightline{LBNL-43560}
\rightline{BNL-HET-99/15}
\rightline{ATL-COM-PHYS-99-017}
\rightline{July, 1999}
\vskip2cm
\centerline{\Large\bf Measurements of Masses in SUGRA Models at 
LHC\footnotemark} 
\footnotetext{This work was supported in part by the Director, Office of
Science, Office of Basic Energy Research, Division of High Energy
Physics of the U.S. Department of Energy under Contracts
DE-AC03-76SF00098 and DE-AC02-98CH10886.}

\bigskip\bigskip
\centerline{\bf Henri Bachacou$^{\bf a,c}$,  Ian Hinchliffe$^{\bf a}$   and Frank E. Paige$^{\bf b}$}
\centerline{\it $^a$ Lawrence Berkeley National Laboratory, Berkeley, CA}
\centerline{\it $^b$ Brookhaven National Laboratory, Upton, NY}
\centerline{\it $^c$ Kungliga Tekniska H\"ogskolan, Stockholm, Sweden} 

\vskip3cm
\centerline{\bf Abstract}
\medskip
\begingroup\narrower\narrower
        This paper presents new measurements  in a  case study of the
minimal SUGRA model with $m_0=100\,\GeV$, $\mhalf=300\,\GeV$, $A_0=0$,
$\tan\beta=2$, and $\sgn\mu=+$ based on four-body distributions from
three-step decays and on minimum masses in such decays. These
measurements allow masses of supersymmetric particles to be determined
without relying on a model.  The feasibility of testing 
slepton universality at the
$\sim0.1\%$ level at high  luminosity is discussed.  In addition, the effect of
enlarging the parameter space of the minimal SUGRA model is discussed.
The direct production of left handed sleptons and the non-observation
of additional structure in the dilepton invariant mass distributions
is shown to provide additional constraints.
\vskip0pt\endgroup

\newpage
\tableofcontents
\newpage

%%%%%%%%%%%%%%%%%%%%%%%%%%%%%%%%%%%%%%%%%%%%%%%%%%%%%%%%%%%%%%%%%%%%%%
\section{Introduction} 
\label{intro}

If SUSY particles exist at the TeV mass scale, they will be produced at
the LHC with large rates, so discovery of their existence will be
straightforward. In minimal SUGRA\cite{SUGRA} and similar models,
however, the decay products of each SUSY particle contain an invisible
lightest SUSY particle (LSP) $\lsp$, so no masses can be reconstructed
directly. Previous studies of SUGRA models have concentrated upon
extracting information from kinematic end points measured in three-body
final states\cite{HPSSY,P1,P3,P4,P5} resulting from decays of the type
$\tchi_2^0 \to \lsp \ell^+\ell^-$ and $\tq_L \to \tchi_2^0 q \to h \lsp
q$. Studies of GMSB\cite{GMSB,HP} models have shown that multi-step
decay chains such as $\tchi_2^0 \to \tell_R^\pm\ell^\mp \to \lsp
\ell^+\ell^- \to \tG \gamma\ell^+\ell^-$ can provide multiple
constraints which can be used to extract masses without fitting to any
underlying model.

In this paper we exploit multi-step decays for SUGRA models,
specifically the decay
\begin{equation}\label{squark}
\tq_L \to \tchi_2^0 q \to \tell^\pm \ell^\mp \to \lsp \ell^+\ell^- q
\end{equation}
for the minimal SUGRA model with the previously studied\cite{HPSSY,P5}
parameters $m_0=100\,\GeV$, $\mhalf=300\,\GeV$, $A_0=300\,\GeV$,
$\tan\beta=2$, and $\sgn\mu=+$. The masses for this point are given in
Table~\ref{mass-table}. As we shall show below, we are able to
reconstruct both upper edges for the $\ell^+\ell^-$, $\ell^+\ell^-q$,
and $\ell^\pm q$ mass distributions and a lower edge for the
$\ell^+\ell^-q$ mass coming from backwards decays of the $\tchi_2^0$ in
the $\tq_L$ rest frame. (The use of analogous upper and lower edges to
reconstruct masses has been extensively discussed for $e^+e^-$
colliders\cite{NLC}.)  These measurements make it possible to
reconstruct all the masses involved in the decay. As part of this
analysis we develop a fitting procedure to make estimates of the
statistical errors of the various measurements that are more
quantitative than previous ones\cite{HPSSY,P5}.

We then illustrate how these and other techniques can be used to
over-constrain the parameters of the minimal SUGRA model and place
constraints on the model itself. As part of this study we will show how
some signals change qualitatively as one varies the relations between
the squark and slepton masses that the minimal SUGRA model predicts. We
allow for the masses of the third generation squarks and sleptons to
vary and for the masses of the sfermions in the {\bf 5} and {\bf 10}
representations of $SU(5)$ to differ. This is the first step in
assessing how well more general supersymmetric models can be constrained
using data from the LHC.

\begin{table}[t]
\caption{Masses of the superpartners, in GeV in the default case
(Point~5) and in two modified cases with $m_5 = 75\,\GeV$ and $m_5 =
125\,\GeV$ described in Section~\ref{vary}.  The first and second
generation of squarks and sleptons are degenerate and so are not listed
separately.
\label{mass-table}} 
\begin{center}
\begin{tabular}{cccccc} \hline \hline 
Superpartner & default & $m_5$ = 75 GeV& $m_5$ = 125 GeV \\ \hline
$\widetilde g$                 &  769 & 769 & 769 \\
$\widetilde \chi_1^\pm$        &  232 & 232 & 232 \\
$\widetilde \chi_2^\pm$        &  523 & 525 & 520 \\
$\widetilde \chi_1^0$          &  122 & 122 & 122 \\
$\widetilde \chi_2^0$          &  233 & 233 & 233 \\
$\widetilde \chi_3^0$          &  502 & 504 & 500 \\
$\widetilde \chi_4^0$          &  526 & 528 & 524 \\
$\widetilde u_L$               &  687 & 687 & 687 \\
$\widetilde u_R$               &  664 & 664 & 664 \\
$\widetilde d_L$               &  690 & 690 & 690 \\
$\widetilde d_R$               &  662 & 659 & 666 \\
$\widetilde t_1$               &  496 & 496 & 495 \\
$\widetilde t_2$               &  706 & 706 & 705 \\
$\widetilde b_1$               &  635 & 635 & 634 \\
$\widetilde b_2$               &  662 & 659 & 666 \\
$\widetilde e_L$               &  239 & 229 & 250 \\
$\widetilde e_R$               &  157 & 157 & 157 \\
$\widetilde \nu_e$             &  230 & 221 & 242 \\
$\widetilde \tau_1$            &  157 & 157 & 157 \\
$\widetilde \tau_2$            &  239 & 230 & 250 \\
$\widetilde \nu_\tau$          &  230 & 220 & 242 \\
$h^0$                          &   95 &  95 &  95 \\
$H^0$                          &  616 & 613 & 618 \\
$A^0$                          &  610 & 608 & 613 \\
$H^\pm$                        &  616 & 613 & 619 \\
\hline \hline
\end{tabular}
\end{center}
\end{table}

The analyses presented here are based large samples of events (the
actual numbers are given below) generated using ISAJET\cite{Isajet} and
RUNDST~5, which implements a simple detector simulation\cite{HPSSY,HP}
including efficiencies representative of the ATLAS detector. Jets were
found using a fixed cone algorithm of size $R=0.4$. Missing energy was
determined including the $\eta$ coverage and a Gaussian approximation to
the energy resolution of the calorimeter. Leptons energy resolutions
were also included (for details see Ref.~\cite{HP}) and an 
appropriate detector efficiency of 90\% was
included. The event selection cuts make the Standard Model background
small compared to the SUSY signal, so that clean SUSY samples can be
studied. In an actual experiment, the cuts are likely to be less severe
so that more signal events will survive. The dominant background with
our cuts is combinatorial background in the interesting SUSY events and
other SUSY events that happen to pass the cuts.
 
This paper contains treats several distinct but closely related topics.
Section~\ref{four-body} describes the extraction of combination of
masses from four-body kinematic limits. Section~\ref{back-edge}
describes the use of a lower edge to determine another combination of
masses, and Section~\ref{ind-mass} combines these measurements to
determine several masses of SUSY particles without relying on a model
fit. Section~\ref{dileperr} discusses the errors that could be achieved
on the dilepton endpoint. We already know\cite{HPSSY,P5} that this error
is quite small, but as precise a measurement as possible is useful as a
test of $e/\mu$ universality. Section~\ref{vary} explores additional
signatures that are relevant to some simple extensions of the minimal
SUGRA model. Section~\ref{fits} reexamines the global fits of parameters
both for the minimal SUGRA model and for its extensions.  Finally, after
some concluding remarks, the appendix describes an unsuccessful attempt
at full reconstruction of SUSY events.

%%%%%%%%%%%%%%%%%%%%%%%%%%%%%%%%%%%%%%%%%%%%%%%%%%%%%%%%%%%%%%%%%%%%%%
\section{Information from four-body decays}
\label{four-body}

In this and the following two sections (\ref{back-edge},
\ref{ind-mass}) and in the Appendix,  the analysis is based on a sample of
$10^6$ SUSY events generated with ISAJET~7.32. This sample corresponds
to approximately 70 fb$^{-1}$ of integrated luminosity. This large
sample is needed so that the the statistical fluctuations shown on the
plots in the these sections corresponds approximately to those expected
in the actual data from one year at the LHC design luminosity. The
Standard Model background is not included on the plots (generating
comparable statistics for this is a prohibitive task) but is known to be
small in the channels\cite{HPSSY,P5} that are used in these sections.
  
Previous work on measurements in SUGRA models relied mainly on end points
measured in three-body final states with one invisible
particle\cite{HPSSY,P1,P3,P4,P5}. It was subsequently found\cite{HP} in
studying GMSB Point G1a that four-body distributions from multi-step
decays such as $\tchi_2^0 \to \tell_R^\pm\ell^\mp \to \lsp\ell^+\ell^-
\to \tG\gamma\ell^+\ell^-$ contain even more information. This method
has considerable generality as we now illustrate by applying it to the
decay chain given in Equation~\ref{squark} above.  The three observed
particles, $\ell^+$, $\ell^-$ and $q$ (which appears as a hadronic jet)
can be used to make several mass distributions: $\ell^+\ell^-q$, $\ell
q$ and $\ell^+\ell^-$.  The last distribution was considered
previously\cite{HPSSY,P5} and has a sharp kinematic end point that
results when the unobserved $\lsp$ momentum is minimized in the rest
frame of $\tell_R^\pm$.

In order to ensure a clean sample of SUSY events the following event
selection was applied.
\begin{itemize}
\item At least four jets with $p_{T,1}>100\,\GeV$ and
$p_{T,2,3,4}>50\,\GeV$, where the jets are numbered in order of
decreasing $p_T$.
\item $\Meff>400\,\GeV$, where $\Meff$ is the scalar sum of the
transverse momenta of the four leading jets and the missing transverse
energy:
$$
\Meff = p_{T,1} + p_{T,2} + p_{T,3} + p_{T,4} + \etmiss\,.
$$
\item  $\etmiss>\max(100\,\GeV,0.2\Meff)$.
\item Two isolated leptons of opposite charge with $p_T>10\,\GeV$,
$|\eta|<2.5$. Isolation being defined so that there is less than 10 GeV
of additional transverse energy in an $R=0.2$ cone centered on the
lepton.
\end{itemize}
With these cuts the Standard Model background is negligible, as can be
seen from Figure~26 of Reference~\citenum{HPSSY}. Thus, Standard Model
backgrounds will not be shown here.

\begin{figure}[t]
\dofig{3.5in}{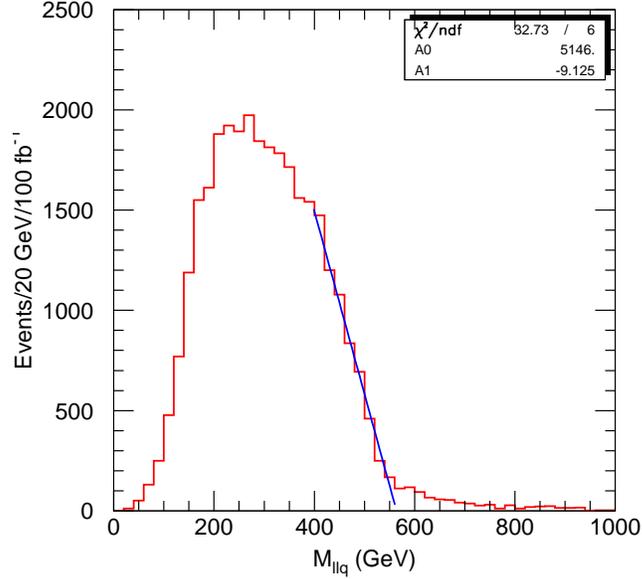}
\caption{Mass distribution for the smaller of the two $\ell^+\ell^-q$
  masses showing a linear fit near the four-body end point. 
\label{c5_mllq0}}  
\end{figure}

It is expected that the two hardest jets will be those coming directly
from $\tq_L \to \tchi_2^0 q$ as a dominant production procees is that
which leads to $\tq_L \tg$ and hence to pairs of $\tq_L$.
 Therefore, the smaller of the two masses
formed by combining the leptons with one of the two highest $p_T$ jets
should be less than the four-body kinematic end point for squark decay,
namely
$$
M_{\ell\ell q}^\rmax = \left[{ \left(M_{\tq_L}^2-M_{\tchi_2^0}^2\right)
\left(M_{\tchi_2^0}^2-M_{\lsp}^2\right) \over M_{\tchi_2^0}^2}
\right]^{1/2} = 552.4\,\GeV\,.
$$
The distribution of the smaller $\ell^+\ell^-q$ mass is plotted in
Figure~\ref{c5_mllq0} with same-flavor lepton pairs weighted positively
and opposite-flavor ones weighted negatively. The $e^+e^- + \mu^+\mu^- -
e^\pm\mu^\mp$ combination cancels all contributions from two independent
decays (assuming $e$-$\mu$ universality) and strongly reduces the
combinatorial background. This distribution should vanish linearly as
the end point is approached.  Figure~\ref{c5_mllq0} also shows a linear
fit near the end point. The extrapolation of this fit gives an end point
of $568.0\,\GeV$, 3.4\% above the nominal value. The distribution itself
is quite linear, so varying the interval over which the fit was made
produces only small changes in the end-point value.

\begin{figure}[t]
\dofig{3.5in}{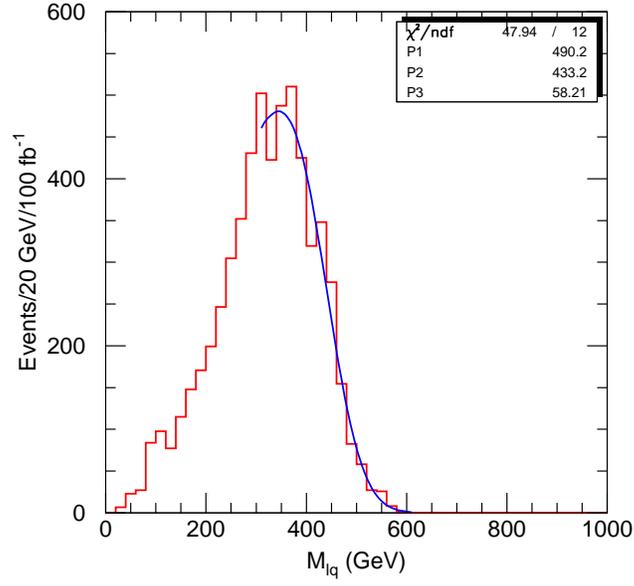}
\caption{Distribution of  the larger of the two $\ell^\pm q$ masses
for $\ell^+\ell^-q$ events in which $M_{\ell\ell q}<600\,\GeV$ and a
fit described in the text. \label{c5_mlq}}
\end{figure}

An additional selection was then made: one $\ell^+\ell^-q$ mass was
required to less than $600\,\GeV$ and the other greater, so that the
assignment of the jet to combine with the lepton pair is unambiguous.
The combination with the smaller mass is then used for further analysis.
The mass distribution of the $\ell^\pm q$ sub-system was then calculated
for each lepton and the selected jet.  If the jet mass is neglected,
then the mass for the jet and the first lepton emitted has an end point
analogous to the $\ell^+\ell^-$ one at
$$
M_{\ell q}^\rmax = \left[{ \left(M_{\tq_L}^2-M_{\tchi_2^0}^2\right)
\left(M_{\tchi_2^0}^2-M_{\ell_R}^2\right) \over M_{\tchi_2^0}^2}
\right]^{1/2}=479.3\,\GeV
$$
In order to make the former structure as clear as possible, the
combination with the larger invariant mass out of the two possible
$\ell^\pm q$ pairings in the $\ell^+\ell^-q$ combination is used in
making Figure~\ref{c5_mlq}.  Again events entered the histogram weighted
by flavor ($+1$ for $e^+e^-$ and $\mu^+\mu^-$ events and $-1$ for
$e^\pm\mu^\mp$) in order to reduce combinatorial background.

There is also an end point where the spectrum vanishes for $M_{\ell q}$
formed using the lepton originating from the last step in the decay
chain, Equation~\ref{squark}, at
$$
M_{\ell q}'^\rmax = \left[{ \left(M_{\tq_L}^2-M_{\tchi_2^0}^2\right)
\left(M_{\ell_R^0}^2-M_{\lsp}^2\right) \over M_{\tchi_2^0}^2}
\right]^{1/2} = 274.5\,\GeV\,.
$$
The structure from this end point is buried under former distribution
and hence is not visible. (In the GMSB case studied previously\cite{HP},
the analogs of both end points were visible.)
 
\begin{figure}[t]
\dofig{3.5in}{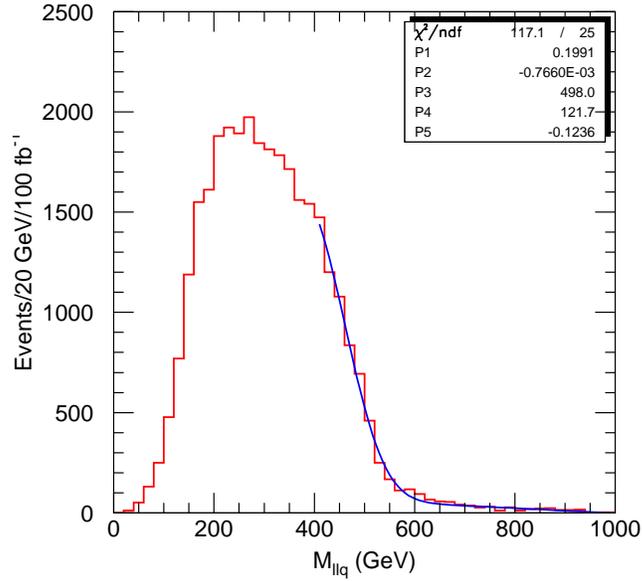}
\caption{Distribution for the smaller of the two $\ell^+\ell^-q$
masses from  Figure~\ref{c5_mllq0};   a Gaussian-smeared fit plus a
linear background described in the text is also shown. 
\label{c5_mllq}} 
\end{figure}

If the resolution were perfect and there were no selection cuts, the
$\ell q$ mass distribution for the lepton emerging from the second
step in the decay chain would be given by
$$
M_{\ell q}^\rmax \sqrt{1+z \over 2}\,dz, \quad -1<z<1
$$
where $M_{\ell q}^{max}$ is the end point given 
above and $z=\cos\theta^*$ is
the cosine of the decay angle of the slepton in its rest frame. In order
to determine whether the selection cuts or resolution are more
responsible for the distortion and to estimate how well this end point
might be measured, this form was smeared with a Gaussian in an attempt
to parametrise the resolution effects.  The fit function is
$$
f(M) = \int_{-1}^{+1}dz\,A\exp\left[{{1\over2\sigma^2} \left(M-M_{\ell
q}^\rmax \sqrt{1+z\over2}\right)^2}\right]
$$
with parameters $A$, $M_{\ell q}^\rmax$, and $\sigma$. The integral was
done numerically using 96-point Gaussian quadrature integration, and the
fit to Figure~\ref{c5_mlq} was made using MINUIT and MINOS\cite{Minos}
inside PAW. The resulting fit, shown in Figure~\ref{c5_mlq}, gives
$M_{\ell{q}}^\rmax = 433.2^{+3.2}_{-3.3}\,\GeV$, which is 9.6\% lower
than the true position and $\sigma=58.2\,\GeV$. It has a reasonable
$\chi^2$, indicating that the cuts do not significantly distort the
shape of the distribution over the fitted range. The shift to lower
values is due primarily to energy lost out of the $R=0.4$ jet cone. If
the analysis is repeated with $R=0.7$ jet cone it is about 3.8\% low.
The resolution is due mainly to the resolution on the jet energy
measurement and is consistent with that expected given the form of our
detector simulation.

The resolution smearing will also shift the position of the $\ell\ell q$
end point.  This distribution was re-fit using the empirical form
$$
f(M) = \int_0^{M_{\ell\ell q}} dz\,
\left({ a_1(M_{\ell\ell q}-z) + a_2(M_{\ell\ell q}-z)^2 }\right)
\exp\left[{ {1\over2\sigma^2} (M-z)^2 }\right]+ b_1 + b_2M 
$$
using the same $\sigma$ obtained above and fitting for $a_i$, $b_i$, and
$M_{\ell\ell q}$. This fit is shown in Figure~\ref{c5_mllq} and gives
$M_{\ell\ell{q}}=498.0^{+7.2}_{-6.4}\,\GeV$, which is 9.8\% low. The
$R=0.7$ jet cone gave a value that was 4.7\% low. In an actual
experiment these shifts due to energy loss out of the cone could be
understood by using detailed comparisons of Monte Carlo simulations with
data and using jets of known energy to set the jet energy
scale\cite{TDR-12}.

The ratio of the $\ell q$ and $\ell\ell q$ end points is independent of
$M_{\tq_L}$:
$$
{M_{\ell q} \over M_{\ell\ell q}} = \sqrt{M_{\tchi_2^0}^2-M_{\ell_R}^2
\over M_{\tchi_2^0}^2-M_{\lsp}^2} = 0.868\,,
$$
compared with a fitted value of $433.2/498.0=0.870$ for the fits
obtained using resolution smearing.  This ratio should be less sensitive
to the jet energy scale and so measured more accurately than the
individual end points. The difference between the fitted and computed
values is very small and the result is stable; repeating the same
analysis using jets defined with an $R=0.7$ cone shifts the individual
fitted edges by $\sim5\%$ but gives 0.877 for the ratio.

%%%%%%%%%%%%%%%%%%%%%%%%%%%%%%%%%%%%%%%%%%%%%%%%%%%%%%%%%%%%%%%%%%%%%%
\section{Lower edges} 
\label{back-edge}

The upper limit of kinematic distributions has been used in the previous
section to extract information. Kinematic distributions can also have
lower limits.  These have been exploited for example in the NLC SUSY
analysis\cite{NLC}. For a process like $e^+e^- \to \tilde\mu^+
\tilde\mu^- \to \mu^+\lsp\mu^-\lsp$, the fixed center of mass energy and
resulting fixed momentum of the $\tilde\mu$ results in a maximum and
minimum energy of the observed muon which can be used to extract both
$\tilde\mu$ and $\lsp$ masses. 

A similar analysis can be used for the process given in
Equation~\ref{squark} at the LHC. The squark mass plays a role analogous
the center of mass energy in the $e^+e^-$ case and a Lorentz invariant
quantity must be used.  For a given value of $z=\cos\theta^*$, the decay
angle of the second lepton in the $\tchi_2^0$ rest frame, the
$\ell^+\ell^-$ mass is determined to be
$$
M_{\ell\ell}^2 = (M_{\ell\ell}^\rmax)^2\,{1+z\over2}\,,
$$
where
$$
M_{\ell\ell}^\rmax = \sqrt{(M_2^2-M_e^2)(M_e^2-M_1^2)\over M_e^2}\,.
$$
There is a corresponding momentum $p_{\ell\ell}$ in the $\tchi_2^0$
rest frame. Thus as a function of $z$ there is a minimum of the
$M_{\ell\ell q}$ mass. For $z=0$ the expression for this minimum
simplifies to
%\begin{eqnarray}
\begin{equation}
\begin{array}{lcl}
(M_{\ell\ell q}^{\rm min})^2 &=& \frac{1}{4 M_2^2 M_e^2} \times \\
&& \Biggl[-M_1^2 M_2^4  + 3 M_1^2 M_2^2 M_e^2 - M_2^4 M_e^2 - M_2^2 M_e^4 
- M_1^2 M_2^2 M_q^2 - \\
&&\quad M_1^2 M_e^2 M_q^2 + 3 M_2^2 M_e^2 M_q^2 - M_e^4 M_q^2 +
(M_2^2-M_q^2)\times \\
&&\quad \sqrt{(M_1^4+M_e^4)(M_2^2 + M_e^2)^2 +
2 M_1^2 M_e^2 (M_2^4 - 6 M_2^2 M_e^2 + M_e^4)}\Biggr]\\
M_{\ell\ell q}^{\rm min}&=& 271.8\,\GeV\nonumber
\end{array}
\label{mllqmin}
\end{equation}
%\end{eqnarray}
where
$$
M_q = M_{\tq_L},\ M_2=M_{\tchi_2^0},\ M_e=M_{\tell_R},\ M_1=M_{\lsp}\,.
$$

\begin{figure}[t]
\dofig{3.5in}{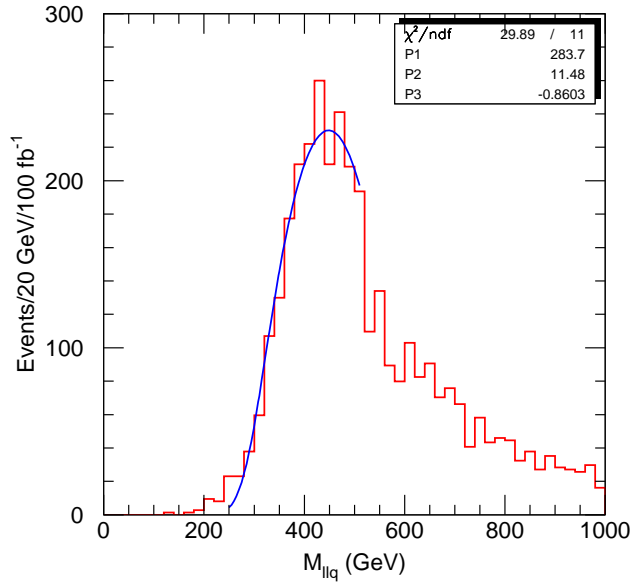}
\caption{Distribution of the larger of the two $\ell\ell q$ masses
after cuts described in the text, showing the lower edge.
\label{c5_100.4_mllqlow}}
\end{figure}

In order to extract $M_{\ell\ell q}^{\rm min}$, events were selected as
before with the additional requirement $M_{\ell\ell}>M_{\ell\ell}^{\rm
max}/\sqrt{2}$, corresponding to $z>0$, and the larger of the two
possible $\ell\ell q$ masses formed by combining the lepton pair with
the two highest $p_T$ jets was chosen.  This distribution is shown in
Figure~\ref{c5_100.4_mllqlow}. The lower edge is not very sharp,
presumably because gluon radiation can carry off energy and so give
masses below the nominal end point.  Nevertheless, the lower edge is
clearly visible and is not obscured by the kinematic cuts.

A fit to the shape including a Gaussian resolution where the width was
allowed to float is unstable; as the fit range was changed the value of
the width changed significantly and was sometimes unreasonably large.
This effect seems to be due to the small tail below the edge.
 Therefore, the Gaussian width was
constrained to be 10\% of the edge value ($M_{\ell\ell q}^{\rm low}$).
This width was used to smear the form
$$
[A(M - M_{\ell\ell q}^{\rm low}) + B(M - M_{\ell\ell q}^{\rm low})^2]
\theta(M -  M_{\ell\ell q}^{\rm low})
$$
which was then fitted to the distribution allowing $A$, $B$ and
$M_{\ell\ell q}^{\rm low}$ to float.  The fit gives
$M_{\ell\ell{q}}^{\rm{low}}=283.7^{+4.4}_{-4.5}\,\GeV$. The $\chi^2$ for
the fit is rather poor (30 for 11 degrees of freedom), mainly because of
the few bins around $200\,\GeV$; a better fit can be obtained by
restricting the range to $300-600\,\GeV$. Because of this, and because
the edge is not very sharp, more study with different choices of the
SUSY parameters is needed to understand the actual error on the fitted
value that could be achieved.  We shall assume conservatively in the
discussion below (see Section~\ref{fits}) that an error of $\pm2\%$ can
be achieved.

\begin{figure}[t]
\dofig{3.5in}{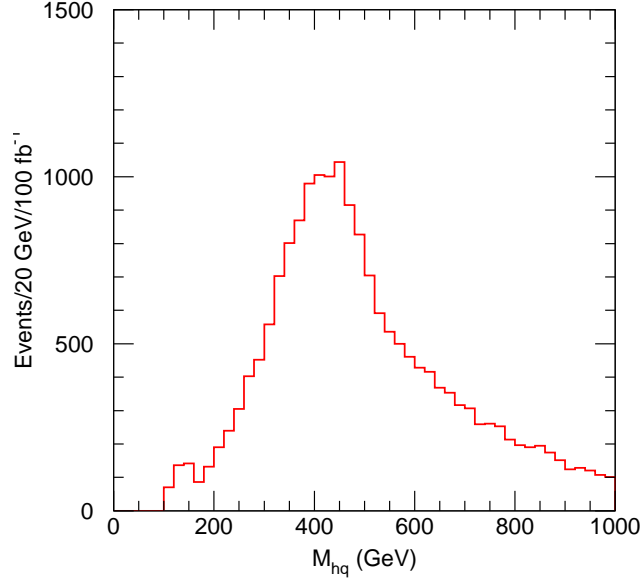}
\caption{Larger of the two $b \bar b q$ jet masses, showing the $hq$
mass threshold. \label{c5_100.4_mhqlow}}
\end{figure}

There is also a minimum value of the $hq$ invariant mass from the decay
chain $\tq_L \to \tchi_2^0 q\to \lsp hq$. If the jet is again treated as
massless, this is given by
\begin{equation}
\begin{array}{lcl}
(M_{hq}^{\rm min})^2 &=& {1 \over 2M_2^2} (M_q^2-M_2^2) \times \\
&& \Bigl[(M_2^2+M_h^2-M_1^2)
-\sqrt{(M_2^2-M_h^2-M_1^2)^2-4M_1^2M_h^2}\Bigr]\\
M_{hq}^{\rm min} &=& 346.5\,\GeV
\end{array}
\label{eq2}
\end{equation}
It has been shown previously that the Higgs decay to $b\overline{b}$
\cite{P5,HPSSY} can be extracted. The following cuts were applied
\begin{itemize}
\item   $\Meff > 400\,\GeV$;
\item   $\etmiss > \max(100\,\GeV,0.2 \Meff)$;
\item   at least four  jets with $p_T > 50\,\GeV$ and  one with $p_{T,1} > 100\,\GeV$;
\item   Transverse sphericity $S_T > 0.2$.
\end{itemize}
In addition, events were selected to have exactly two $b$ jets with
$p_{T,b} > 25\,\GeV$, $70<M_{bb}<110\,\GeV$, and the larger of the two
masses formed by combining this pair with either of the two hardest jets
was selected, since at least one of the two should be greater than the
minimum mass. This distribution is shown in
Figure~\ref{c5_100.4_mhqlow}. While there is a threshold in roughly the
right place, it is not very distinct.  This probably is due to a
combination of resolution for this multijet system and of the
substantial combinatorial background under the Higgs peak. A side-band
subtraction and careful jet energy calibration might be able to clean up
the distribution. We will assume below an error of $\pm5\%$; this is
substantially larger than the $\ell\ell{q}$ error and will only add a
very weak additional constraint. It would be important to study this
further in cases where the decay to sleptons is not available such as
``Point~1'' and ``Point~2'' of Refs.~\citenum{HPSSY} and \citenum{P1}.

%%%%%%%%%%%%%%%%%%%%%%%%%%%%%%%%%%%%%%%%%%%%%%%%%%%%%%%%%%%%%%%%%%%%%%
\section{Model-independent masses}
\label{ind-mass}

In this section we discuss how the measurements in the previous section
and those discussed previously \cite{HPSSY,P5} can be used to determine
the masses of the SUSY particles without reference to the underlying
SUGRA model. The identification of the decay chains is needed but these
are based on much weaker assumptions.  We have a number of measurements
all related to the process $\tq_L \to \tchi_2^0 \to \tell_R \to \lsp$:

\begin{figure}[t]
\dofigs{3.2in}{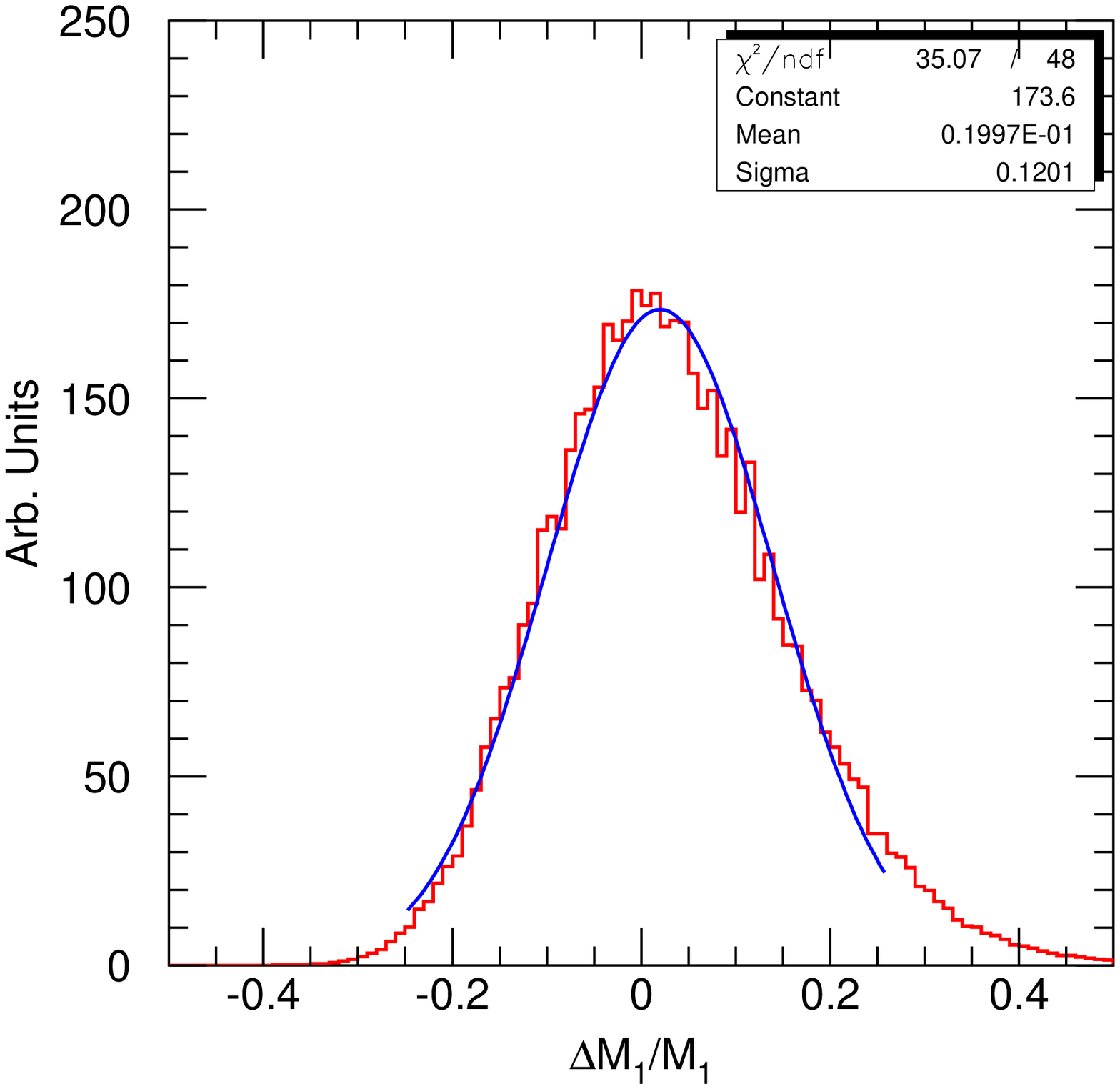}{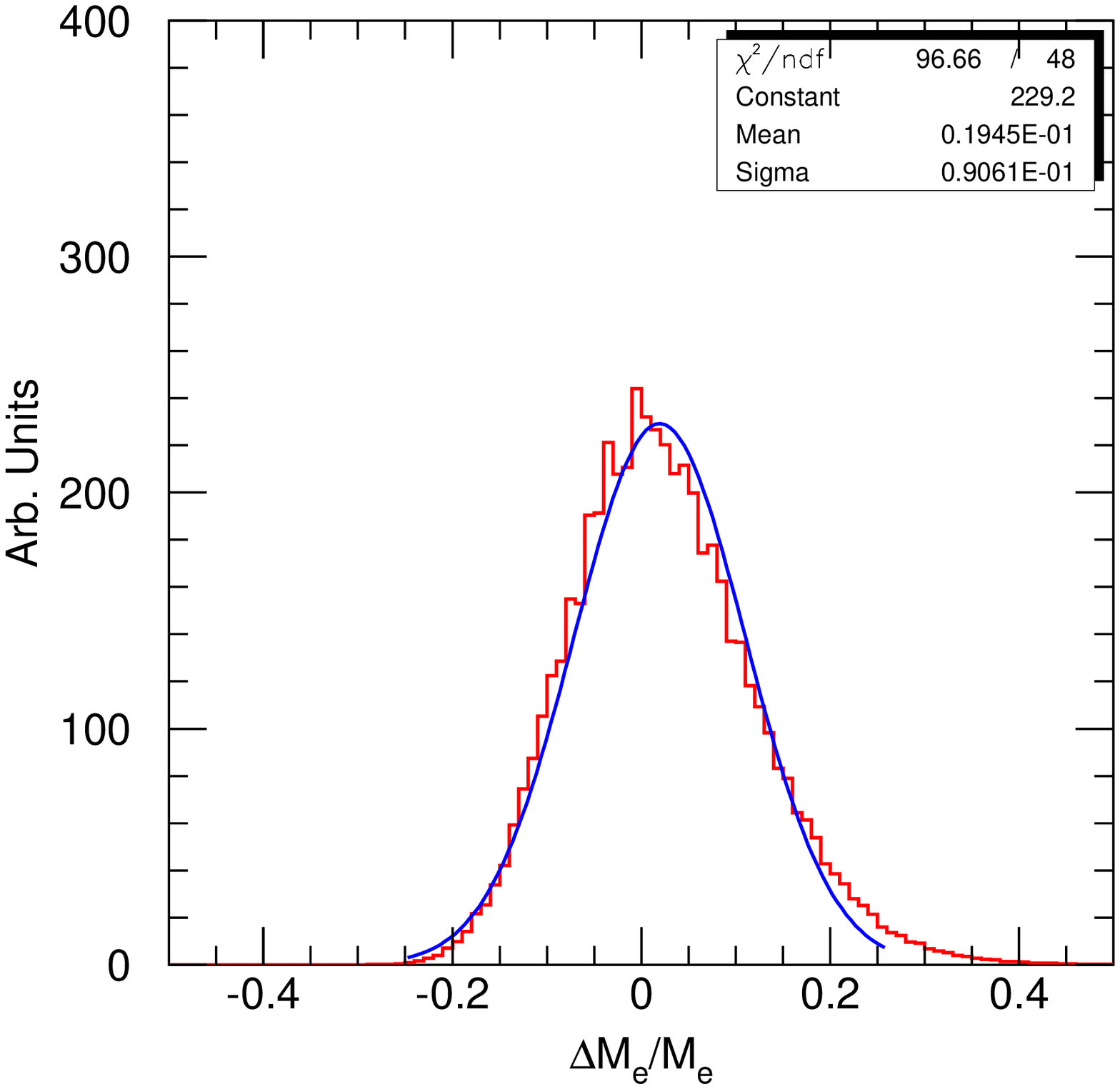}
\dofigs{3.2in}{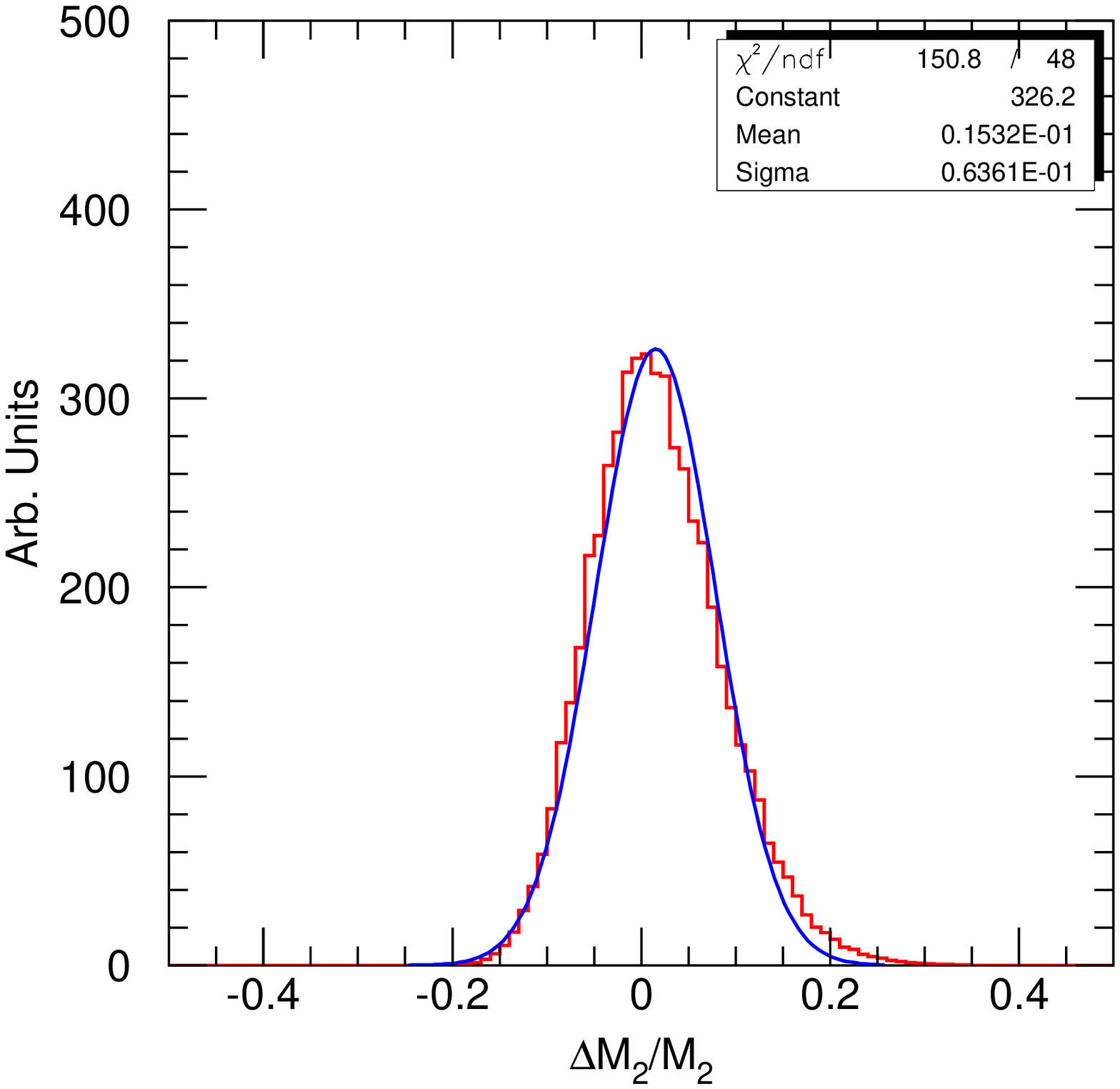}{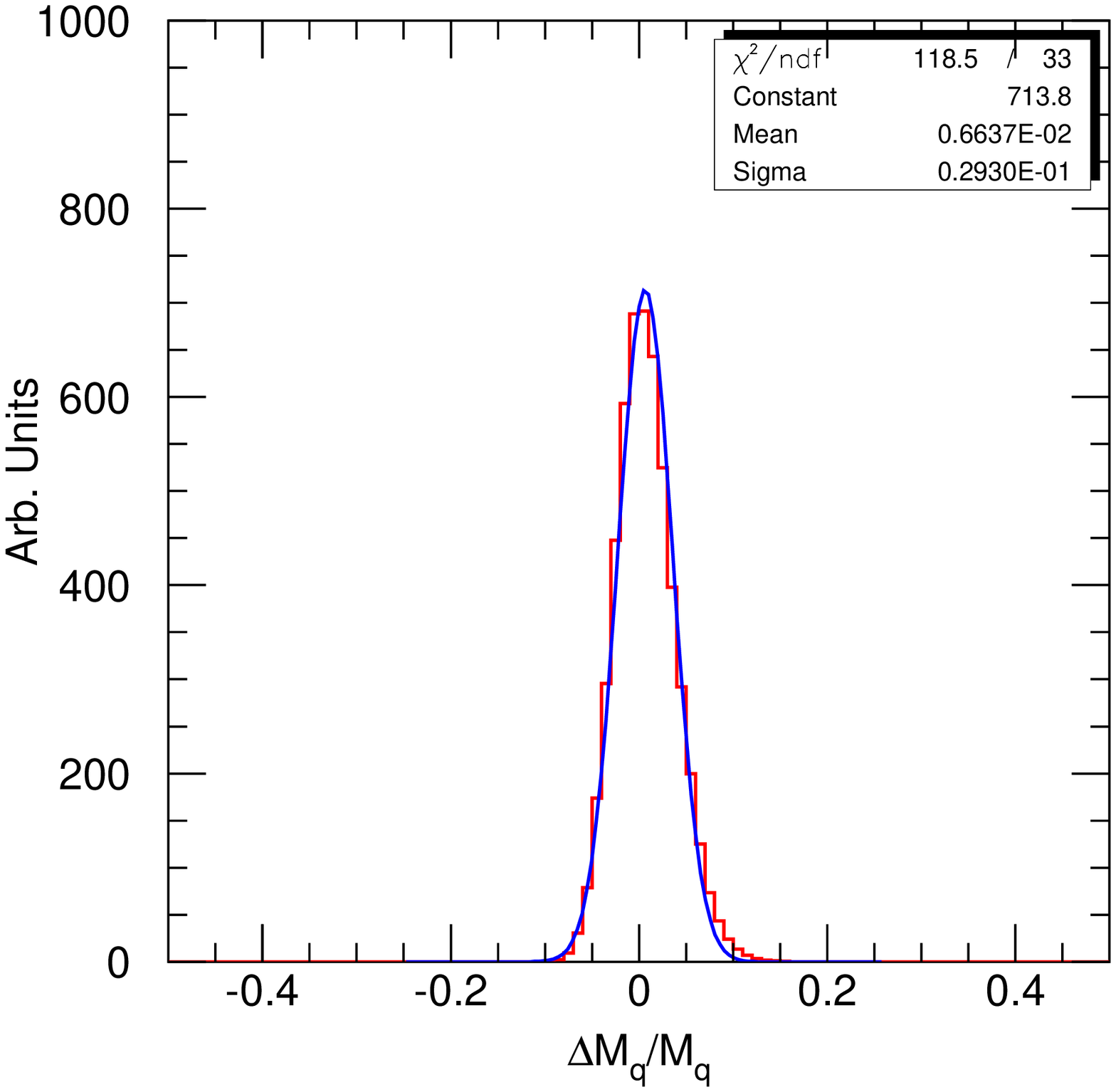}
\caption{Distribution of the $\lsp$, $\tell_R$, $\tchi_2^0$, and $\tq_L$
masses satisfying all constraints discussed in the text. The fitted
widths are about $\pm12\%$, $\pm9\%$, $\pm6\%$, and $\pm3\%$
respectively. \label{c5_back_all}}
\end{figure}

\begin{itemize}
\item   $M_{\ell\ell}^\rmax = \left[{(M_2^2-M_{\tell_R}^2)(M_{\tell_R}^2-M_1^2)
\over M_{\ell_R}^2}\right]^{1/2} = 108.9\pm0.11\,\GeV$ (see Ref~\cite{HPSSY})
\item   $M_{\ell\ell q}^\rmax = \left[{
(M_{\tq_L}^2-M_{\tchi_2^0}^2)
(M_{\tchi_2^0}^2-M_{\lsp}^2) \over M_{\tchi_2^0}^2}
\right]^{1/2} = 552.4\pm5.5\,\GeV$
\item   $M_{\ell q}^\rmax = \left[{(M_{\tq_L}^2-M_{\tchi_2^0}^2)
(M_{\tchi_2^0}^2-M_{\tell_R}^2) \over M_{\tchi_2^0}^2}
\right]^{1/2} = 479.3\pm5.5\,\GeV$
\item   $M_{\ell\ell q}^{\rm min} \hbox{\ (Equation~\ref{mllqmin})} = 
271.8\pm5.4\,\GeV$
\end{itemize} 
There are also two measurements related to $\tchi_2^0 \to \lsp h$:
\begin{itemize}
\item The maximum $hq$ mass (see Reference~\citenum{HPSSY})
$$
(M_{hq}^{\rm max})^2 = M_h^2 
+ \left(M_{\tq}^2 - M_{\tchi_2^0}^2\right) 
\left[{M_{\tchi_2^0}^2 + M_h^2 - M_{\lsp}^2 + 
\sqrt{(M_{\tchi_2^0}^2 - M_h^2 - M_{\lsp}^2)^2 -4M_h^2
M_{\lsp}^2}} \over 2 M_{\tchi_2^0}^2 \right]\,.
$$  
which has the value $M_{hq}^{\rm max}=522.6\pm5.2\,\GeV$.
\item The minimum $hq$ mass $M_{hq}^{\rm min} 
\hbox{\ (Equation~\ref{eq2})} = 346.5\pm17.3\,\GeV$
\end{itemize}
A large error is assigned to the $hq$ lower edge because it is
not sharp and there is a lot of background (see above). 

\begin{figure}[t]
\dofig{3.5in}{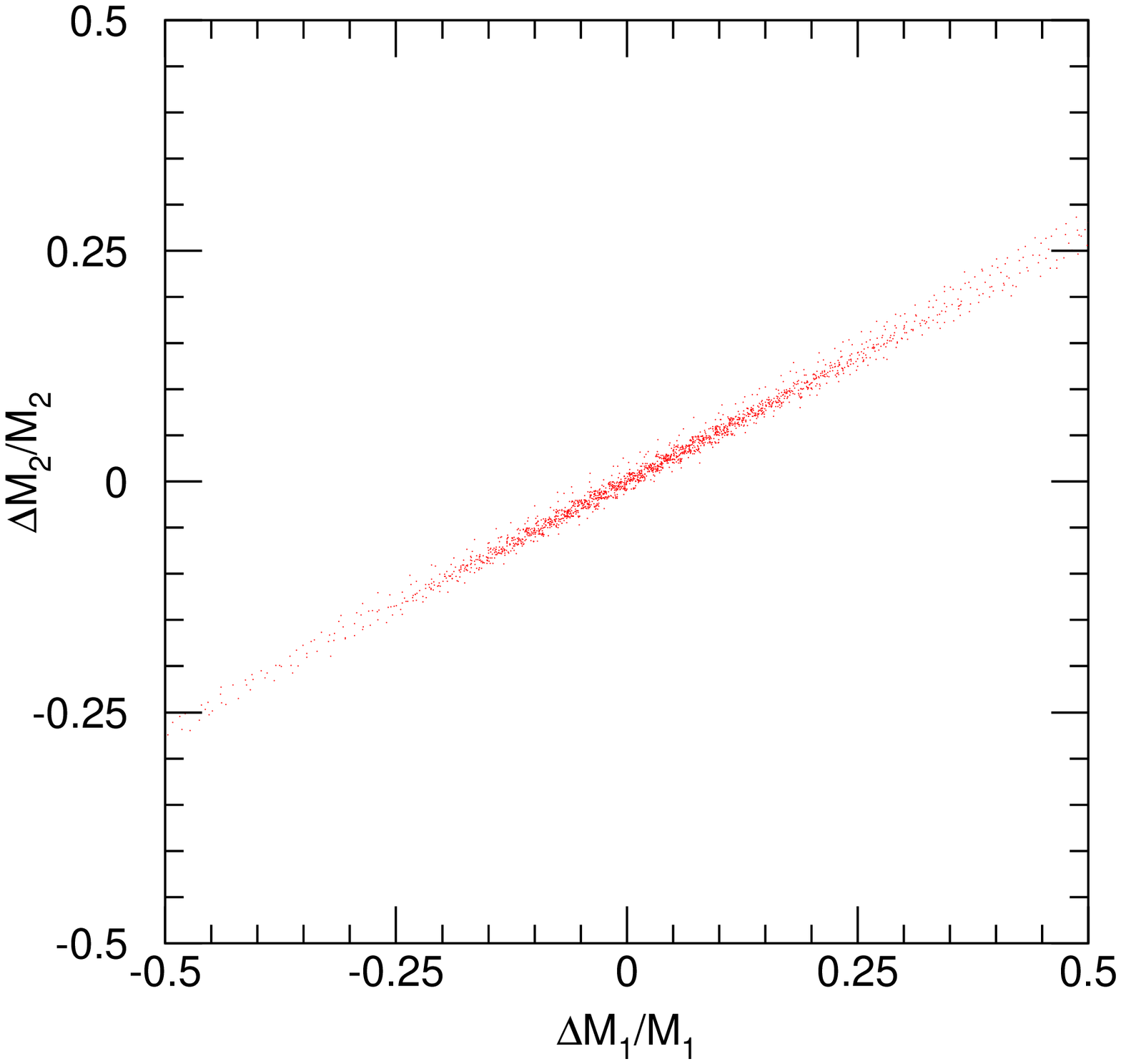}
\caption{Scatter plot of $M_{\tchi_2^0}$ vs.\ $M_{\lsp}$ solutions
satisfying all constraints discussed in the text.
\label{c5_back_m2vm1}}
\end{figure}

\begin{figure}[t]
\dofig{3.5in}{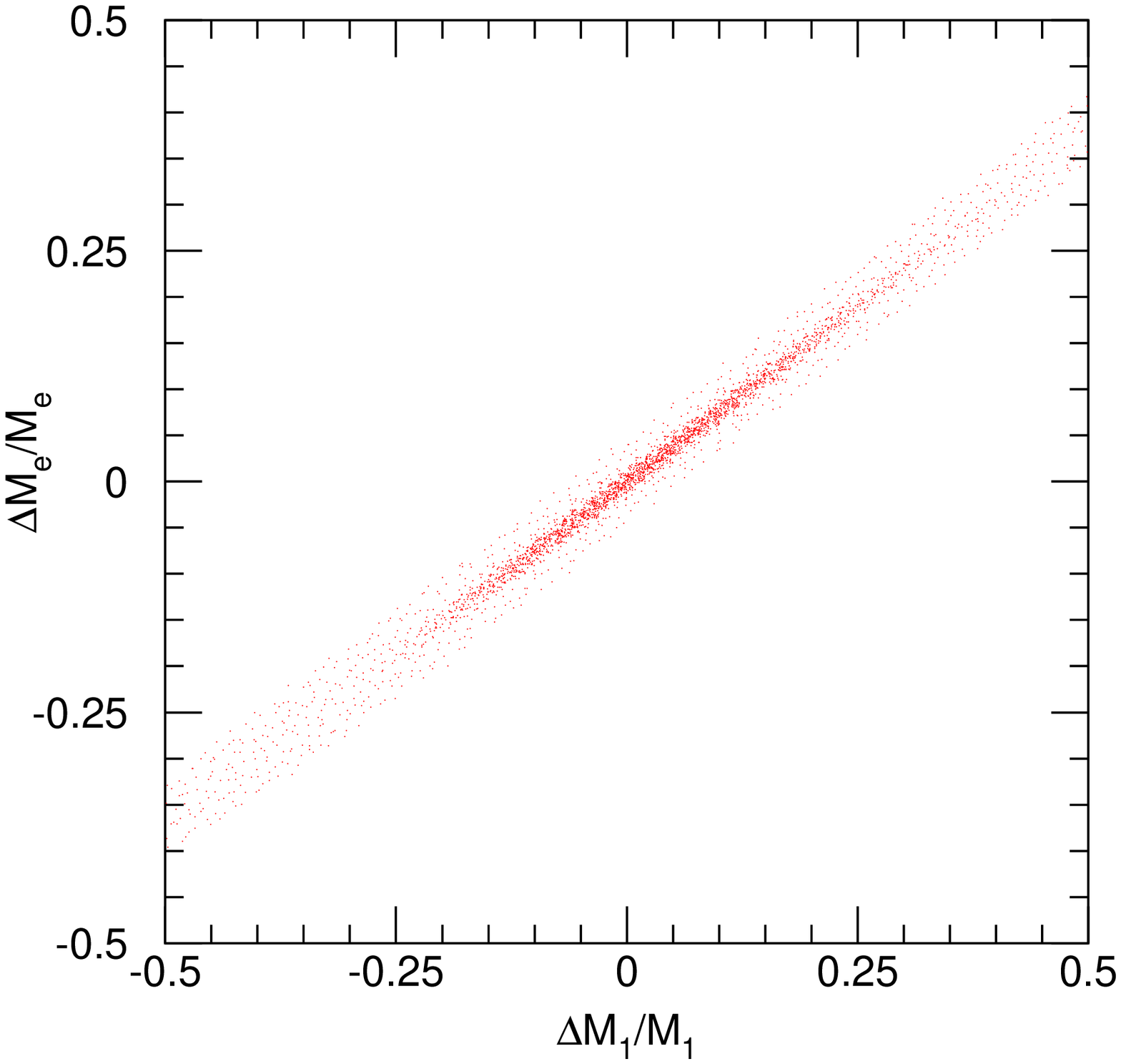}
\caption{Scatter plot of $M_{\tell_R}$ vs.\ $M_{\lsp}$ solutions
satisfying all constraints discussed in the text.
\label{c5_back_mevm1}}
\end{figure}

If the lower edges discussed in the previous section are ignored, there
are four measurements and four unknown masses masses $M_{\tilde{q_L}}$,
$M_{\tchi_2}$, $M_{\tilde{\ell_R}}$ and $M_{\lsp}$. Nevertheless, for
the errors assumed there is a one-parameter family of solutions labeled
by $M_{\lsp}$, with small uncertainties in the other masses for a fixed
value of $M_{\lsp}$. This remains true even if the errors are
substantially reduced.

If the lower edges are included, then all four masses can be determined.
The errors were estimated numerically as follows.  The $\tq_L$,
$\tchi_2^0$, and $\tell_R$ masses were generated uniformly within
$\pm50\%$ of their nominal values, and the $\lsp$ mass was calculated
using the dilepton edge ($M_{\ell\ell}$), which has a much smaller error
than the other measurements. The $\chi^2$ for the remaining measurements
was calculated, and the point was assigned a probability of
$\exp(-\chi^2/2)$. The resulting distributions are shown in
Figure~\ref{c5_back_all}. The resulting errors range from $\pm12\%$ for
the mass of $\lsp$ to $\pm3\%$ for the mass of $\tq_L$. If the error on
$M_{\ell\ell q}^{\rm min}$ were reduced to $\pm1\%$, as might be
possible with a more careful understanding of the systematics,
 the error on $M_{\lsp}$ would be reduced
to $\pm7.5\%$. The errors are highly correlated as can be seen from
Figures~\ref{c5_back_m2vm1}--\ref{c5_back_mevm1} which show the
scatter plots of $M_{\tchi_2^0}$ {\it vs. } $M_{\lsp}$ and $M_{\tell_R}$
vs.\ $M_{\lsp}$. Of course, the errors on the masses are much poorer than
those that arise from a fit within the SUGRA model (see
Section~\ref{fits} and Ref.~\cite{HPSSY}), but they do not involve any
model assumptions.

%%%%%%%%%%%%%%%%%%%%%%%%%%%%%%%%%%%%%%%%%%%%%%%%%%%%%%%%%%%%%%%%%%%%%%
\section{Dilepton measurement errors}
\label{dileperr}

In order to provide significant constraints on model parameters and
tests of the underlying model a number of measurements with {\it
comparable} errors are needed.  In general, measuring one combination
of SUSY masses very precisely is not particularly useful if the other
combinations involve jets and so are only measured with an accuracy of
several percent. An important exception is the decay $\tchi_2^0 \to
\tell_R^\pm \ell^\mp \to \lsp \ell^+\ell^-$, which has an end point at 
\begin{equation}
M_{\ell\ell}^{\rm max} = M_{\tchi_2^0}
\sqrt{1 - {M_{\tell_R}^2 \over M_{\tchi_2^0}^2}}
\sqrt{1 - {M_{\lsp}^2 \over M_{\tell_R}^2}}\,.
\label{mllmax}
\end{equation}
A difference in the end points for $e^+e^-$ and $\mu^+\mu^-$ would
directly indicate a difference in the $\tilde e_R$ and $\tilde\mu_R$
masses, which obviously an important issue for testing models that
purport to understand flavor physics.  This decay is generally allowed
in SUGRA models which give cosmologically interesting cold dark
matter\cite{BB}, such as the one discussed here. It is also common in
GMSB models since the $\tilde\ell_R$ has only $U(1)$ couplings and tends
to be light. The derivative of the end-point with respect to $M_{\tell_R}$
vanishes at the geometric mean of the $\lsp$ and $\tchi_2^0$ masses but
in general is of order one; for the masses in the case studied here,
$$
{d M_{\ell\ell}^{\rm max} \over d M_{\tell_R}} = 0.478\,.
$$

\begin{figure}[t]
\dofig{3.5in}{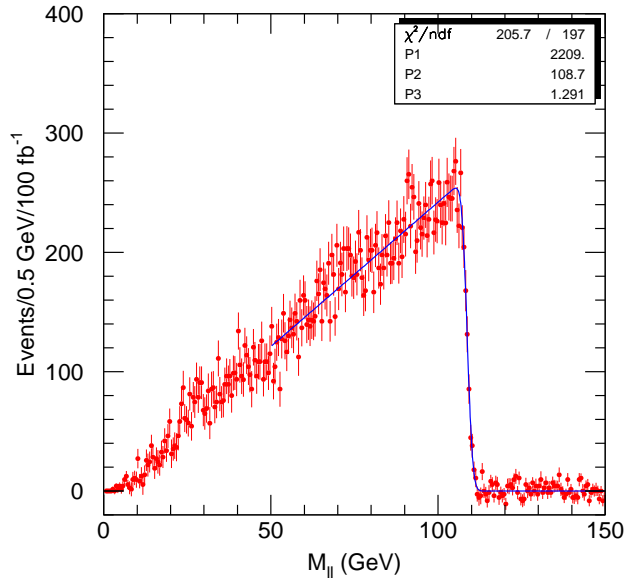}
\caption{$\ell^+\ell^-$ mass distribution showing the  $\chi^2$
MINUIT fit using PAW. \label{c5_mllfit}}
\end{figure}

\begin{figure}[t]
\dofig{3.5in}{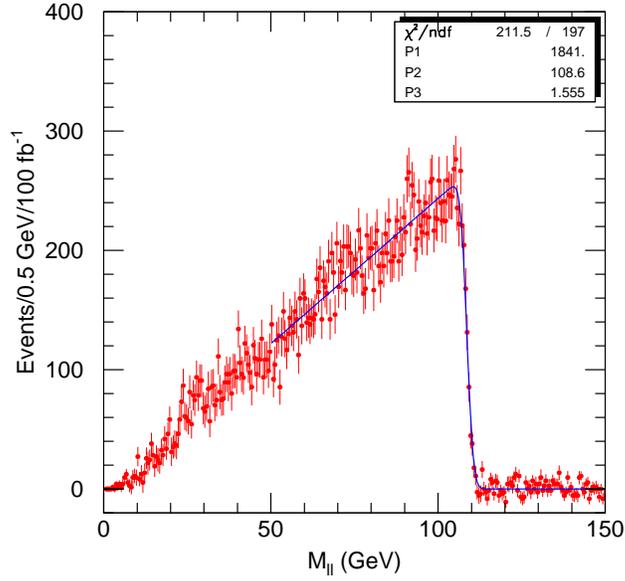}
\caption{$\ell^+\ell^-$ mass distribution showing the   maximum
likelihood MINUIT fit using PAW.\label{c5_mllfit2}}
\end{figure}

The same sample of $10^6$ SUSY events was used to estimate how well
such an edge might be measured with full LHC luminosity. In the
absence of cuts, the mass distribution should be given by the same
formula as discussed in Section~\ref{four-body}, namely
$$
(M_{\ell\ell}^{\rm max})^2 {1+z \over 2} dz
$$
with $z$ uniformly distributed. This form was smeared with a Gaussian
using numerical integration as in Section~\ref{four-body}.
Figures~\ref{c5_mllfit} and \ref{c5_mllfit2} show the resulting fits
using MINUIT with either the $\chi^2$ or the maximum likelihood
method; the parameters are the overall normalization, the end point,
and the Gaussian width. The fitted end points with errors from
MINOS\cite{Minos} are $108.71{+0.087\atop-0.088}\,\GeV$ and
$108.60{+0.065\atop-0.060}\,\GeV$ respectively. The fits are consistent
with each other, but neither quite agrees within errors with the
expected end point at $108.92\,\GeV$. The statistical errors are
slightly better than the systematic errors expected from the lepton
energy scale\cite{TDR-12} and are comparable to the errors on the $W$
mass expected to be achieved ultimately at the Tevatron\cite{Tev-W} and
LEP\cite{LEP-W}.

\begin{figure}[t]
\dofig{3.5in}{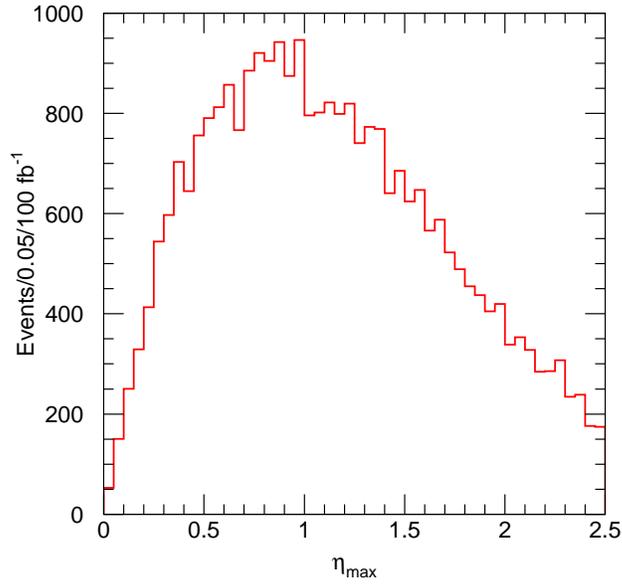}
\caption{Maximum $|\eta_\ell|$ for dilepton events.
\label{c5_etamax}}
\end{figure}

The maximum $|\eta_\ell|$ for the either of the two leptons is plotted
in Figure~\ref{c5_etamax} and peaks around $\eta_\rmax=1$. Thus both the
barrel and the endcap regions of the detector are important. If precise electron and
muon measurements were available only in the barrel, about half the
events would be lost.

Clearly a lot more work is needed to understand how to calibrate the
detector and to extract information at this level of accuracy. In
particular, the discrepancy between the fitted and calculated end points
even in this highly idealized simulation needs to be understood,
presumably by studying many different samples of SUSY events. It seems
clear, however, that statistical errors below about 0.1\% may be
achievable, especially if the masses were a bit lower so that the
$\tchi_2^0 \to \lsp h$ decay were absent.

%%%%%%%%%%%%%%%%%%%%%%%%%%%%%%%%%%%%%%%%%%%%%%%%%%%%%%%%%%%%%%%%%%%%%%
\section{Non-minimal SUGRA models}
\label{vary}

The case studied so far assumes that the scalar masses are all equal at
the GUT scale, an assumption that is rather restrictive and may not be
valid. By studying variations in this assumption, we can try to estimate
howl the various LHC signals are modified and how well this
assumption could be tested. We shall show that qualitatively new signals
emerge in our preliminary study of non-universal SUGRA (NUSUGRA) models
which is carried out using two kinds of NUSUGRA models closely related
to the SUGRA case discussed above.

\subsection{Variations of masses with $SU(5)$ representations}

We vary the scalar masses at the unification scale by assuming that
squarks and sleptons which are in the {\bf 10} of SU(5) have a common
scalar mass $m_{10}$ while those that are in the {\bf 5} of SU(5) have a
mass $m_5$. Here, $m_{10}$ is kept at its nominal value of $100\,\GeV$
while $m_5$ is shifted. Two points have been studied in detail, namely
$m_5 = 75\,\GeV$ and $m_5 = 125\,\GeV$.  The masses of the superpartners
at these points are given in Table~\ref{mass-table}.  Squark masses are
almost insensitive to these changes in $m_5$; the shifts are much
smaller than the errors that were obtained in Section~\ref{ind-mass}
because $m_{1/2}$ plays a dominant role in these masses via the strong
coupling of squarks. The significant changes take place in the slepton
spectrum, in particular in the masses of $\tilde{\ell_L}$. Samples of 200000
events were simulated for each of the new cases; $10^6$ Standard Model
background events were also used to ensure that the cuts are effective
in disposing of it.

\subsubsection{The $m_5 < 100\,\GeV$ case}
\label{m5-down}

In the benchmark case, $m_5=m_{10}=100\,\GeV$, the decay sequence
$\tilde\chi_2^0 \ra \tilde\ell_R^\pm \ell^\mp \ra \lsp \ell^+\ell^-$ is
allowed, giving a very clear signature: the lepton pair invariant mass
distribution presents a sharp edge near the kinematic limit,
Equation~\ref{mllmax}.  For modified points with $m_5 < 85 \,\GeV$, the
left handed slepton becomes lighter than $\tilde\chi_2^0$ thus the decay
sequence $\tilde\chi_2^0 \ra \tilde\ell_L^\pm \ell^\mp \ra \lsp
\ell^+\ell^-$ is also allowed, giving rise to a second edge at
$$
M_{\ell\ell}'^\rmax = M_{\tilde\chi_2^0} 
\sqrt{1-{M_{\tilde\ell_{L}}^2 \over M_{\tilde\chi_2^0}^2}}
\sqrt{1-{M_{\lsp}^2 \over M_{\tilde\ell_{L}}^2}}  \approx 34.9\,\GeV\,
\hbox{\ for\ }\,m_5 = 75\, \GeV\,.
$$
The position of the edge is sensitive to $M_{\tilde\ell_{L}}$ and thus
to $m_5$. As there is less available phase space in this decay than in
the decay to $\ell_R$, these leptons are softer and it is necessary to
lower the cuts as much as possible to ensure good acceptance. ATLAS
expects to be able to detect muons down to $p_T = 5\,\GeV$, so the
following selection cuts are applied:

\begin{figure}[t]
\dofig{3.5in}{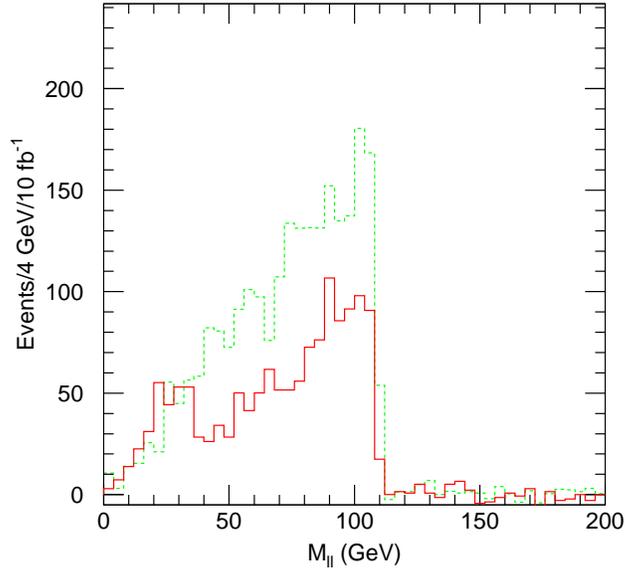}
\caption{$M_{\ell\ell}$ distribution after the cuts at the
$m_5=m_{10}=100\,\GeV$ point (dashed line) and modified point with $m_5
= 75\,\GeV$ (solid line). \label{mllnu75}}
\end{figure}

\begin{figure}[t]
\dofig{3.5in}{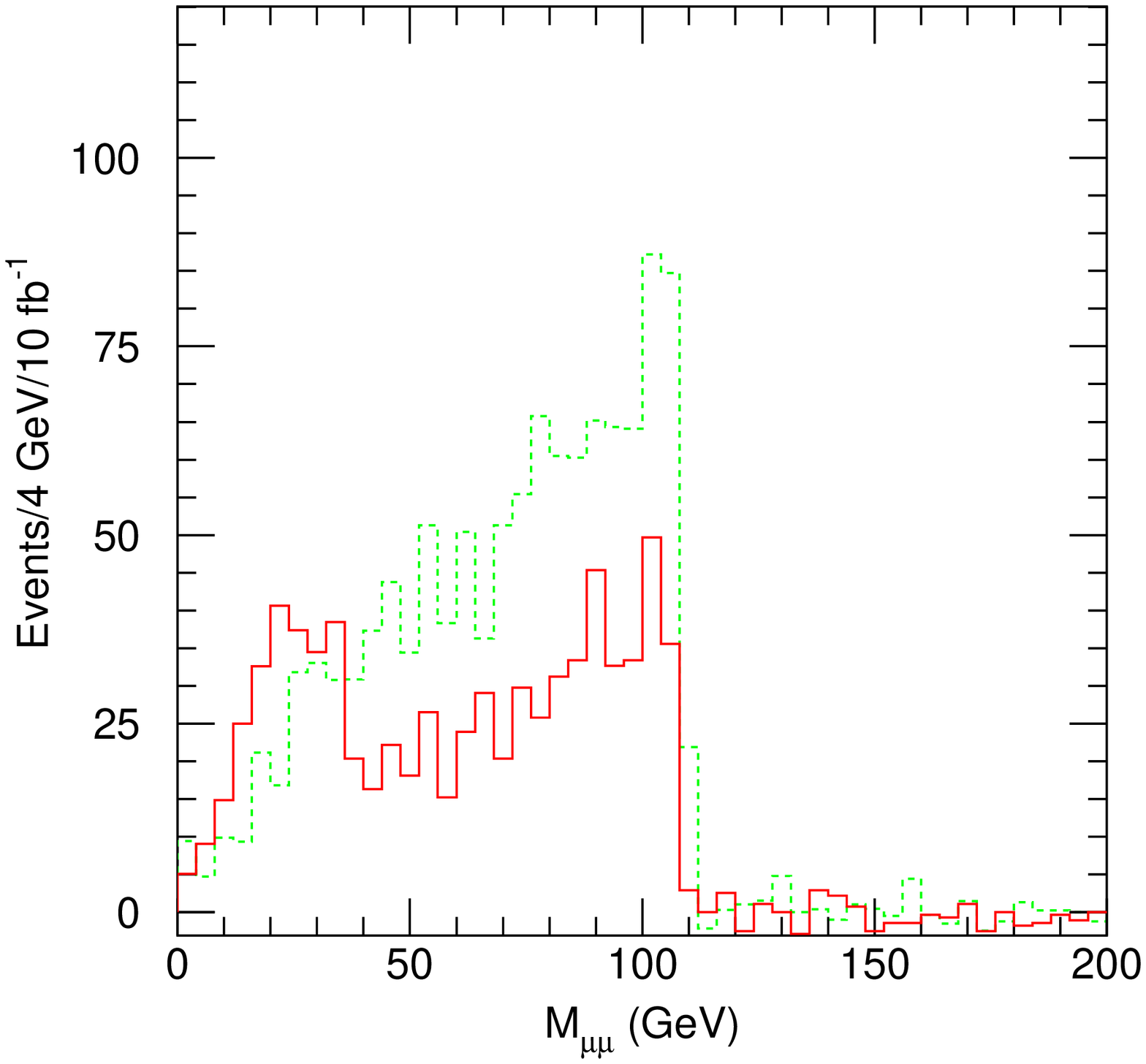}
\caption{$M_{\mu\mu}$ distribution after the cuts at the
$m_5=m_{10}=100\,\GeV$ point (dashed line) and modified point with $m_5
= 75\,\GeV$ (solid line). \label{mmumunu75}}
\end{figure}

\begin{figure}[t]
\dofig{3.5in}{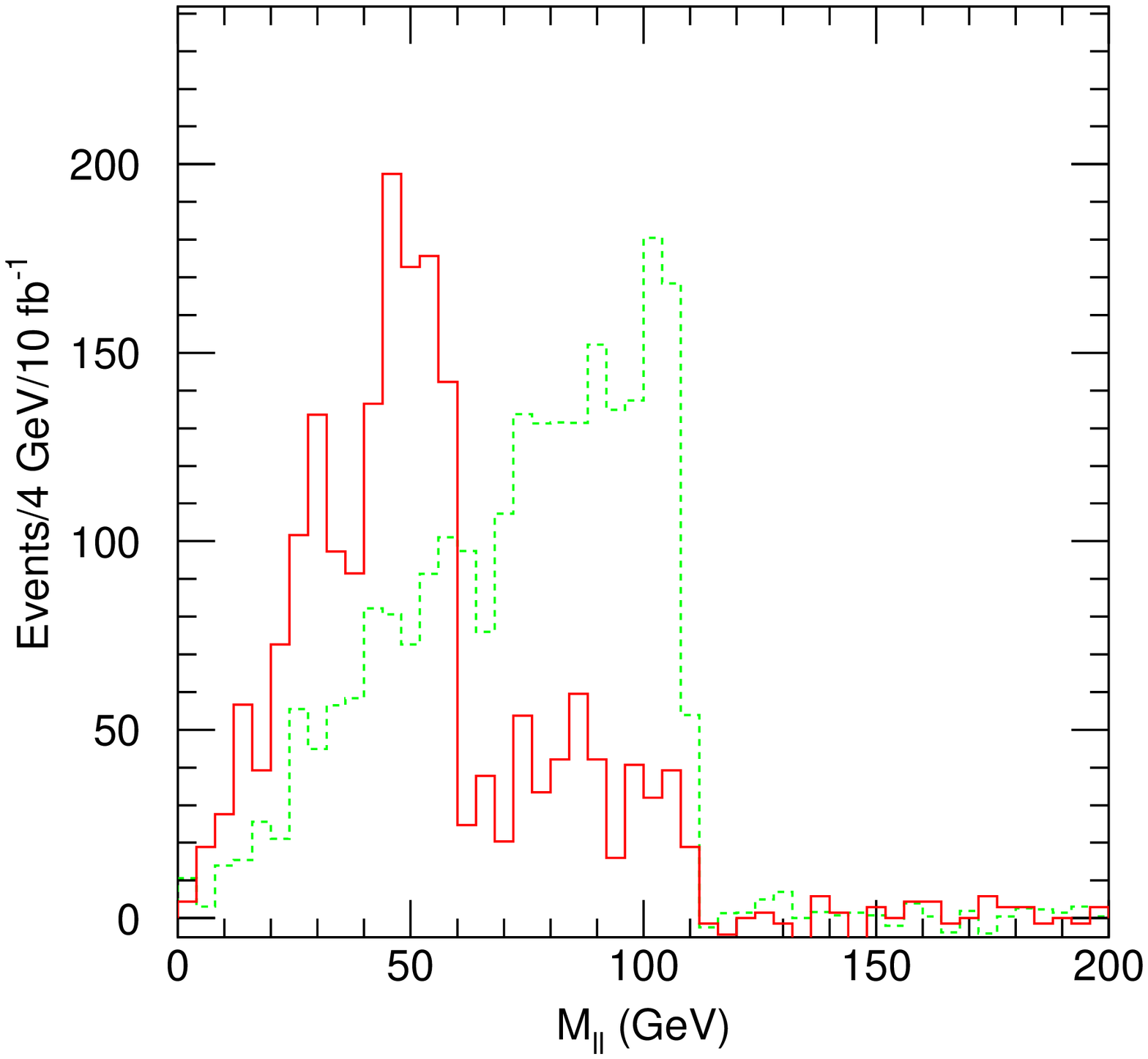}
\caption{$M_{\ell\ell}$ distribution after the cuts at the
$m_5=m_{10}=100\,\GeV$ point (dashed line) and modified point with $m_5
= 50\,\GeV$ (solid line). \label{mllnu50}}
\end{figure}

\begin{figure}[t]
\dofig{3.5in}{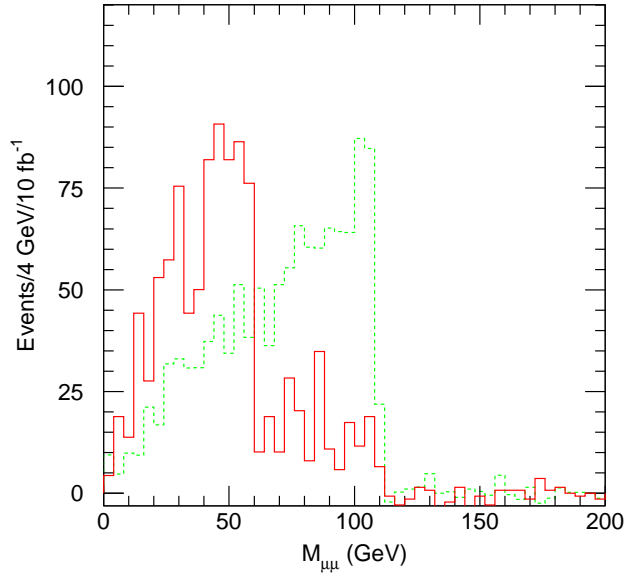}
\caption{$M_{\mu\mu}$ distribution after the cuts at the
$m_5=m_{10}=100\,\GeV$ point (dashed line) and modified point with $m_5
= 50\,\GeV$ (solid line). \label{mmumunu50}}
\end{figure}

\begin{itemize}
\item   $\Meff > 800\,\GeV$;
\item   $\etmiss > 0.2\Meff$;
\item   at least one $R=0.4$ jet with $p_T > 100\,\GeV$;
\item   $\ell^+\ell^-$ pair with $|\eta_\ell| < 2.5$, $p_{T,e}> 10\,\GeV$, and
$p_{T,\mu} > 5\,\GeV$;
\item   $\ell$ isolation cut: $E_T < 10\,\GeV$ in $R=0.2$ around the leptons;
\item   Transverse sphericity $S_T > 0.2$.
\end{itemize}

Figures~\ref{mllnu75} and \ref{mllnu50} show the lepton pair invariant
mass for the benchmark case and the modified cases with $m_5 = 75\,\GeV$
and $m_5 = 50\,\GeV$. Figures~\ref{mmumunu75} and \ref{mmumunu50} show
the muon pair invariant mass for the same cases.  The edge at low
$M_{\ell\ell}$ for $m_5 = 75\,\GeV$ is clearer in the muon case due to
the increased acceptance at low $p_T$. The presence
of two structures with comparable rates enables one to deduce the
presence of two decay chains and to measure the two end points.  Notice
that as $m_5$ is reduced to 50 GeV, the higher mass structure is
becoming weaker because the $\tilde\chi_2^0\ell\tilde{\ell_L}$ coupling
is larger than the $\tilde\chi_2^0\ell\tilde{\ell_R}$ one.  As $m_5$
increases, $M_{\tilde\ell_{L}}$ increases, and the branching ratio for
$\tilde\chi_2^0 \ra \ell \tilde\ell_{L}$ vanishes quickly around
$m_5$ = 80 GeV. Hence, $m_5 = 75\,\GeV$ is about the upper limit where
one can distinguish from the benchmark case using this channel.

\subsubsection{The $m_5 > 100 \,\GeV$ case}
\label{lept}

In this case there is no visible effect on the $\ell^+\ell^-$
distribution or any of the other distributions studied in the previous
sections. Production of $\tell_L$ via the decay of strongly interacting
sparticles is small, so one must rely on direct production via the
Drell-Yan process. It is possible\cite{P5} to extract a signal for
Drell-Yan production with $\tell_L \to \lsp\ell$ by requiring two
isolated leptons, missing energy, and no jets, but the slepton mass can
only be inferred from the rate and kinematic distributions. The jet
veto is essential to elliminate events with leptons arising from the
decays of squarks and gluinos.

If $m_5$ is somewhat larger than $100\,\GeV$, the $\ell_L$ becomes heavy
enough that the decay $\tell_L \to \tchi_2^0 \ell$ is allowed. Then the
decay chain
$$
\arraycolsep=0pt % Stupid LaTeX
\begin{array}{llll}
\tilde\ell_L &  & + \;\;\;\;\;\;& \tilde\ell_L \\
\downarrow &&& \downarrow \\
 \lsp +\ \ell^\pm &&& \tchi_2^0 + \ell^\mp\\
&&& \downarrow \\
&&& \tilde\ell_R +  \ell^{\prime\pm} \\
&&& \downarrow  \\
&&& \lsp + \ell^{\prime\mp} \\
\end{array}
$$
results in a final state with four isolated leptons. The signature is
two same flavor opposite charge (SFOC) lepton pairs and no jet activity.
In order to select these events, the following cuts were applied:
\begin{itemize}
\item   no jet with $p_T > 40\,\GeV$ and $|\eta| < 5$;
\item   at least 4 leptons with $p_T > 10\, \GeV$  and  $|\eta| < 2.5$
forming 2 SFOC pairs;
\item   the invariant mass of at least one of the pairs is less than 109 GeV
  (so that it is a candidate to arise from the decay of  
$\tchi_2^0$);
\item   $\ell$ isolation cut: $E_T < 10\,\GeV$ in $R=0.2$ around the lepton.
\end{itemize}

The invariant mass of the three leptons coming from the same left-handed
slepton provides information about its mass. The three leptons were
selected as follows:
\begin{itemize}
\item   a pair with invariant mass smaller than 109 GeV is assumed to come 
from the $\tchi_2^0$ decay;
\item   the remaining lepton with lowest $p_T$ is assumed to come from the 
same left handed slepton as the pair.
\end{itemize}
The invariant mass of the trilepton system is then computed.  As the
production rate is small, high luminosity is needed and an integrated
luminosity of $300\,\fbi$, representing the ultimate that can be
achieved at LHC, was assumed.  The $M_{\ell\ell\ell}$ invariant mass
should have an upper limit given by:
$$
M_{\ell\ell\ell}^{\rm max} 
=\sqrt{
\left(1-{M_{\tilde\ell_{R}}^2 \over M_{\tilde\chi_2^0}^2}\right)
\left(M_{\tilde\ell_{L}}^2\left( 1-{M_{\tilde\chi_2^0}^2 \over 
M_{\tilde\ell_{L}}^2}\right)
+M_{\tilde\chi_2^0}^2\left(1-{M_{\lsp}^2 \over M_{\tilde\ell_{R}}^2}\right)
\right)
}
$$
which is approximately 128.0 GeV for  $m_5 = 125\,\GeV$.

\begin{figure}[t]
\dofig{3.5in}{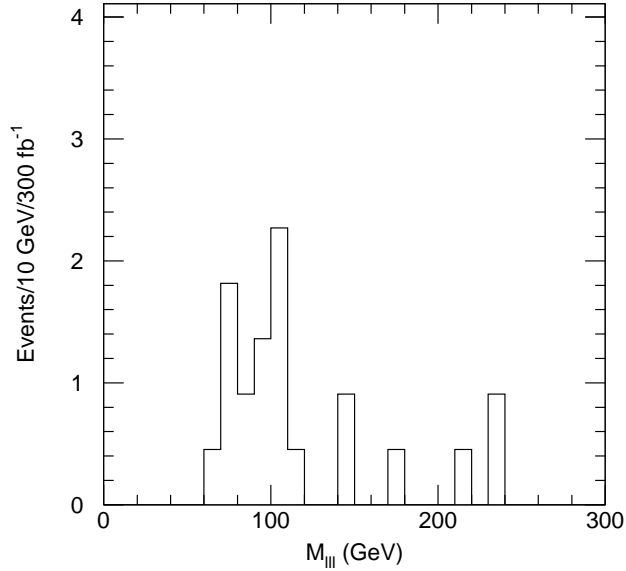}
\vskip-14pt
\caption{$M_{\ell\ell\ell}$ distribution after the cuts for the
$m_5=m_{10}=100\,\GeV$ point. Note that the number of events is very
small.
\label{mlllpoint5}}
\end{figure}

\begin{figure}[t]
\dofig{3.5in}{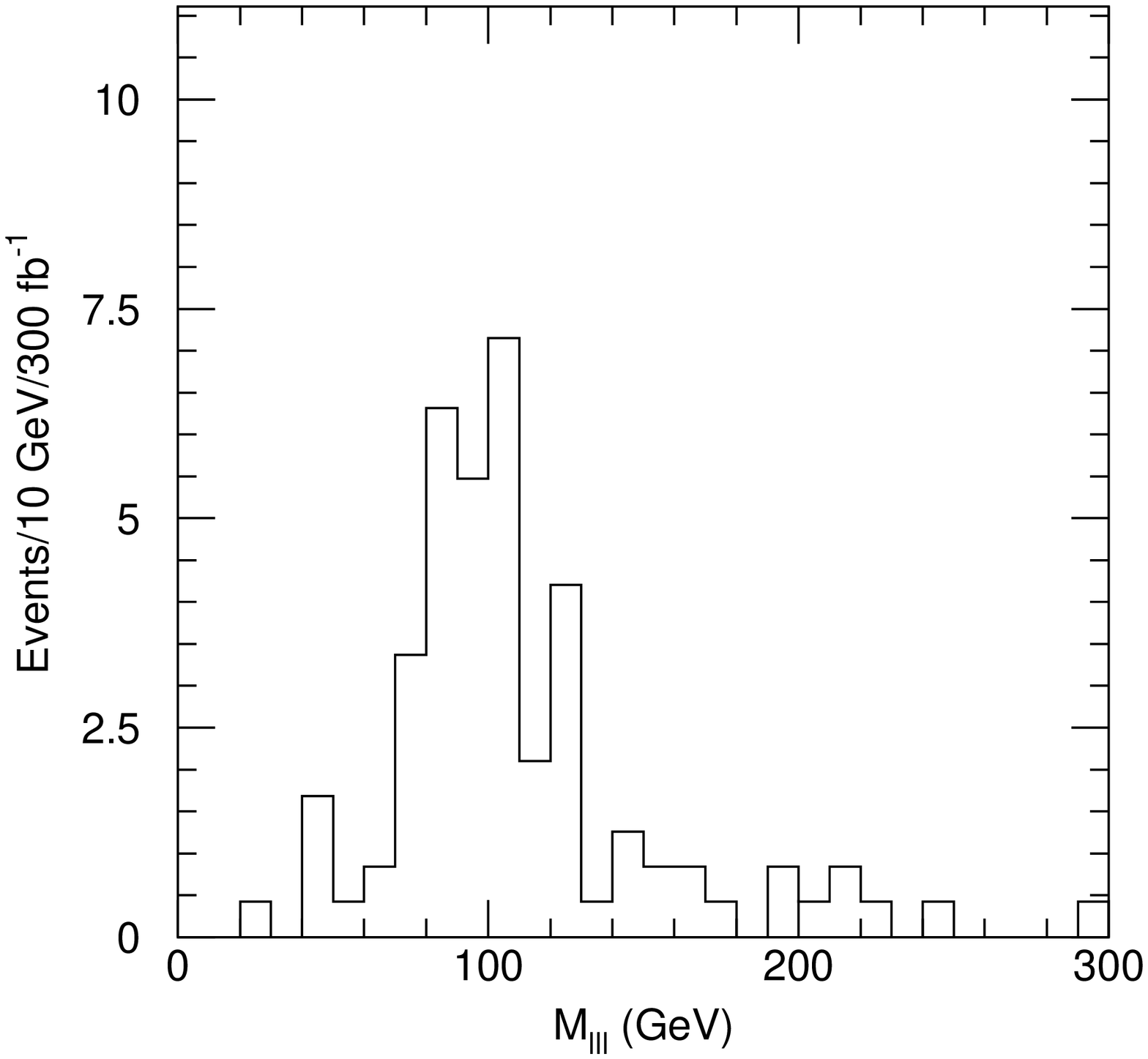}
\vskip-14pt
\caption{$M_{\ell\ell\ell}$ distribution after the cuts for the modified
point with $m_5$ = 115 GeV. \label{mlllnu115}}
\end{figure}

\begin{figure}[t]
\dofig{3.5in}{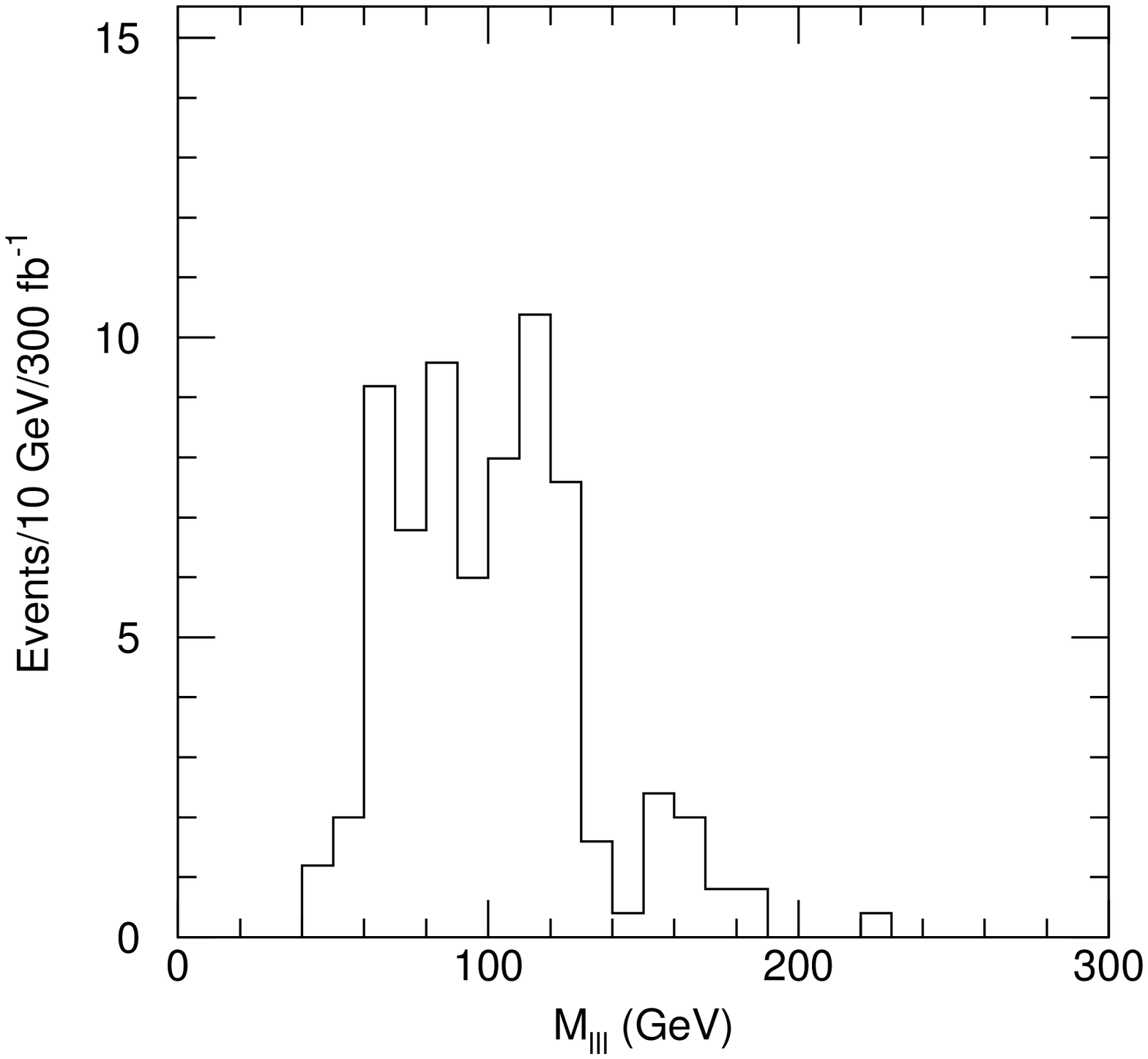}
\vskip-14pt
\caption{$M_{\ell\ell\ell}$ distribution after the cuts for the modified
point with $m_5$ = 125 GeV. \label{mlllnu125}}
\end{figure}

Figures~\ref{mlllpoint5},~\ref{mlllnu115}~and~\ref{mlllnu125} show the
tri-lepton invariant mass respectively at point 5, and the cases with
$m_5 = 115\,\GeV$, and $m_5$ = 125 GeV.  As $M_{\tilde\ell_{L}}$
increases, the $\tilde\ell_{L} \ra \tilde \chi_2^0 \ell$ branching
ratio increases (2\% for point 5, 10\% for $m_5 = 125\,\GeV$).  A
clear signal appears for $m_5 = 115\,\GeV$. We estimate that for
masses larger than this, the position of the edge can be measured with
a precision of 3 GeV and is very sensitive to $m_5$.  As $m_5$
increases further, the production rate falls off and the signal
disappears for $m_5$ above 250 GeV. 

\subsection{Variations of masses for the third generation}
\label{thisrd}

Here we investigate the possibility that $m_0$ for the third
generation squarks and sleptons is different from that for the first
two generations. As in the $m_5$ cases, the greatest sensitivity to
this change is in the slepton sector, so we are forced to consider the
detection of final states containing taus.

\begin{figure}[t]
\dofig{3.5in}{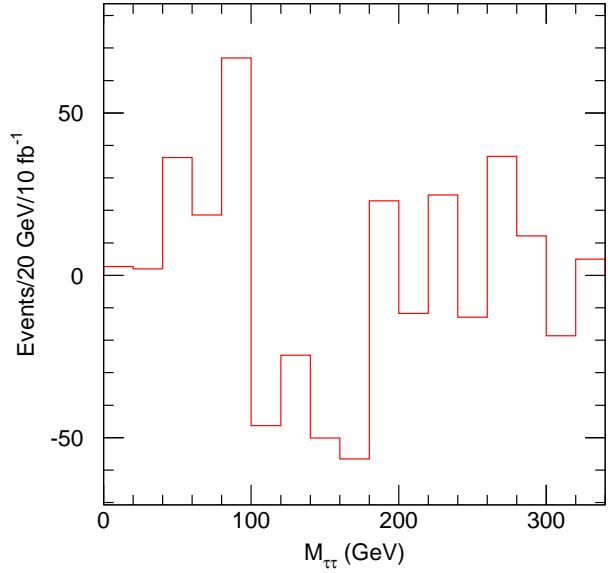}
\caption{ $M_{\tau^+\tau^-}$ distribution at point 5 after the cuts
and after subtraction of same sign pairs. \label{mtt}}
\end{figure}

In another study\cite{point6}, the use of hadronic tau decays was
illustrated. It was shown how the $\tau\tau$ invariant mass distribution
could be inferred from the observed decay products and the kinematic end
point extracted. This method enables $m_{\tilde{\tau_R}}$ to be
constrained. If this method could be exploited in the case of interest
here, slepton universality could be tested with great accuracy. The
method fails for several reasons all of which are related to the  observable
event rate and background.
\begin{itemize}
\item The signal is less clear because the channels $\tilde\chi_2^0
\ra h^0 \tilde \chi_1^0$ and $\tilde \chi_2^0 \ra \tilde \ell_R \ell$,
$\ell$ being $e$ or $\mu$, are open here, whereas the $\tilde \chi_2^0$
decays only to staus for the case studied in Ref.~\citenum{point6}. The
decay to $h$ also generates $\tau\tau$ final states at a comparable rate
and distorts the shape of the distribution.  
\item Because the gluinos and squarks are heavier, the total SUSY cross section
for the cases discussed here is smaller than in the case studied in Ref~\cite{point6},
resulting in smaller event samples for the same integrated luminosity.
\end{itemize}
Figure~\ref{mtt} shows the $\tau\tau$ invariant mass reconstructed
from the visible decay products for our benchmark case after the
following selection cuts are applied:
\begin{itemize}
\item   $\etmiss > 0.2\Meff\,\mbox{or} > 100\,\GeV$;
\item   at least one $R=0.4$ jet with $p_T > 100\,\GeV$;
\item   at least three additional $R=0.4$ jets with $p_T > 50\,\GeV$;
\item   two jets identified as hadronic tau decays according to the
  methods described in Ref.~\citenum{point6}.
\end{itemize}
The distribution is shown for the subtracted combination $\tau^+\tau^-
-\tau^\pm\tau^\pm$ as this is reduces the background from jets that
are misidentified as taus.

\begin{figure}[t]
\dofig{3.5in}{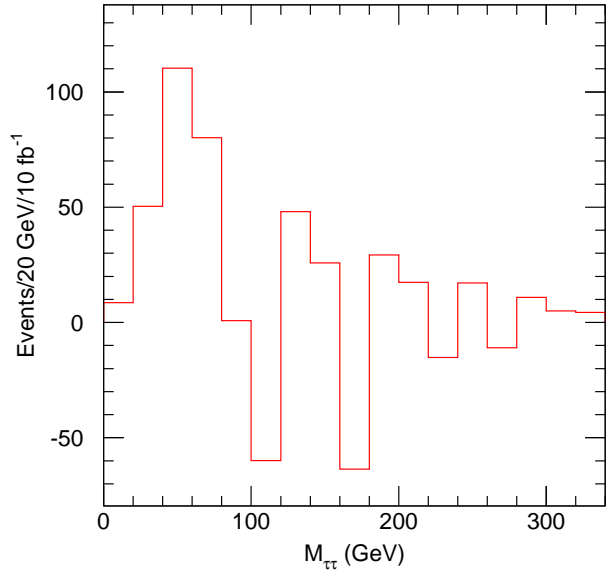}
\caption{ $M_{\tau^+\tau^-}$ distribution at modified point with $tan
\beta = 5$ (other parameters at their nominal point 5 value) after the
cuts and after subtraction of same sign pairs. \label{mtttan5}}
\end{figure}

A sample of 100000 events have been generated; the plot is normalized to
$10\,\fbi$, which would correspond to corresponding to $\sim$~230000
events so the statistical fluctuations are somewhat bigger than they
would be in the actual experiment.  The signal, of about 40 events in
the bin at 50 GeV, is completely buried by the statistical fluctuations. Even several
years of data taking at low luminosity would not be enough to reduce
them to a satisfying level.  LHC high luminosity would give enough
statistics, but the $M_{\tau\tau}$ reconstruction method used here is
based on full simulation\cite{ihtau} and has not been proven to be
viable at high luminosity; additional pile-up events could compromise
it.

To emphasize the deterioration of the signal due to the $\tilde\chi_2^0
\ra h^0 \tilde \chi_1^0$ channel, a model with $\tan\beta$ shifted from
2.1 to 5 (the other SUGRA parameters remaining at their nominal values)
has been studied. Since $h^0$ is now heavier, this decay channel is
closed and the rate of tau pair production is 3 times greater.
Figure~\ref{mtttan5} shows that the excess of $\tau^+\tau^-$ pairs now
becomes visible; there are approximately 120 signal events in the
plot concentrated in the bin at 50 GeV.

An alternative method of extracting evidence for excess $\tau$-pair
production based on leptonic final states is now illustrated. The method
is similar to that of Ref~\cite{denegri}. In order to illustrate its
sensitivity, we have studied models in which the SUGRA parameters remain
at their nominal values except that the third generation squark and
slepton masses (namely $t_l$, $b_r$, $t_r$, $L_l$, $L_r$) are set equal
to $m_{3^{rd}}$ at GUT scale.  The masses of the relevant superpartners
for several values of $m_{3^{rd}}$ are given in Table~\ref{tau-mass},
from which it can be seen that the largest effect is in the stau masses:
as $m_{3^{rd}}$ increases the taus masses rise and the branching ratio
for $\tilde\chi_2^0 \ra \tau \tilde{\tau_1}$ is reduced. The channel
closes for $m_{3^{rd}}> 200\,\GeV$.

\begin{table}[t]
\caption{Masses of the relevant superpartners, in GeV, in the default
case and at modified points with $m_{3^{rd}}$ = 30, 70, 150 and 200
GeV.  The first and second generation of squarks and sleptons are
degenerate and so are not listed separately.
\label{tau-mass}} 

\begin{center}
\begin{tabular}{cccccc} \hline \hline 
Sparticle &  default &  $m_{3^{rd}}$ =  30 GeV&  
$m_{3^{rd}}$ =  70 GeV & $m_{3^{rd}}$  = 150 GeV& $m_{3^{rd}}$  = 200 GeV  \\
 \hline
$\widetilde g$          &769  & 769 & 769 & 769&769 \\
$\widetilde u_l $ &687 & 687 & 687 & 687& 687\\ 
$\widetilde u_r$ &664 & 664 & 664 & 664& 664\\ 
$\widetilde t_1$ &496 & 491 & 493 & 501& 510\\ 
$\widetilde t_2$ &706 & 701 & 704 & 712& 721\\ 
$\widetilde \chi_1^\pm$ &231 & 232 & 232 & 233& 234\\ 
$\widetilde \chi_2^\pm$& 514 & 514 & 515 & 533&549 \\ 
$\widetilde\chi_1^0$   &122  & 122 & 122 & 122&122 \\ 
$\widetilde \chi_2^0$  &232  & 233 & 233 & 234&235 \\ 
$\widetilde e_L$       &239  & 239 & 239 & 239&239 \\ 
$\widetilde e_R$       &157  & 157 & 157 & 157&157 \\ 
$\widetilde \nu_e$     &230  & 230 & 230 & 230&230 \\ 
$\widetilde \tau_1$    &124  & 157 & 140 & 193&234 \\  
$\widetilde \tau_2$    &219  & 239 & 228 & 264&295 \\
$\widetilde \nu_\tau$  & 209 & 230 & 219 & 256&288 \\
\hline \hline
\end{tabular}
\end{center}
\end{table}

Samples of 200000 SUSY events were generated in each case shown in the
table, (except for $m_{3^{rd}} = 30\,\GeV$ and $200\,\GeV$~where 100000
events were generated).  The Standard Model background has been added
using a 1 million event sample. The only significant  background after 
applying our selections is from
$t\overline{t}$ events. In
order to select SUSY events with slepton pair, and reject Standard Model
background, the following cuts have been applied:

\begin{itemize}
\item   $\Meff > 500\,\GeV$;
\item   $\etmiss > \mbox{max}(0.2\Meff, 250\,\GeV)$;
\item   at least  one $R=0.4$ jet with $p_T > 100\,\GeV$;
\item   at least  four $R=0.4$ jets with $p_T > 50\,\GeV$;
\item   $\ell \ell$ pair with $p_{T,\ell}> 10\,\GeV$, $|\eta_\ell| <
2.5$, $\ell$ being electron or muon;
\item   $\ell$ isolation cut: $E_T < 10\,\GeV$ in $R=0.2$;
\end{itemize}

After these cuts the signal exceeds the Standard Model background by
more than a factor of ten; the background is mainly due to top quark
pair production.  Lepton pairs in the signal events arise mainly from
two processes.
\begin{itemize}
\item The decay of $\chi_2^0$, via $\tilde\chi_2^0 \ra \tilde\ell^\pm
\ell^\mp \ra \lsp \ell^+\ell^-$ where $\tilde\ell$ is $\tilde e_R$,
$\tilde \mu_R$, $\tilde \tau_1$, or $\tilde \tau_2$; $\ell$ is the
corresponding lepton.  This channel produces exclusively oppositely
(OC) pairs, of the same flavor (SFOC) in case of selectrons or smuons,
and both same (SFOC) and opposite flavor (OFOC) pairs in case of
leptons from resulting from leptonic tau decays.
\item The decay of a $\tilde \chi_1^\pm$ pair, produced in the decays
of gluinos:  even if the $\tilde g \ra \tilde \chi_1^\pm qq$ branching
ratio is small, this channel is important since it can produce same
charge (SC) pairs, as gluinos are Majorana fermions. The flavors of
the leptons are uncorrelated.  The main $\tilde \chi_1^\pm$ decay
modes are:  $\tilde \chi_1^\pm \ra \lsp W^\pm \ra \lsp \ell^\pm
\nu_\ell ({\bar\nu}_\ell)$ $\tilde \chi_1^\pm \ra \lsp \ell^\pm
\nu_\ell$ $\tilde \chi_1^\pm \ra \tilde \nu_\ell \ell^\pm$, $\ell$
being $e$, $\mu$ or $\tau$.
\end{itemize}

A violation of $e,\mu,\tau$ universality will be revealed by comparing
the number of events containing same flavor lepton pair (SF), with the
number of events containing opposite flavor lepton pair (OF). Define
$$r_{OC} = \frac{e^+e^- + \mu^+\mu^-}{e^\pm\mu^\mp}$$ This ratio
decreases as the branching ratio for $\tilde\chi_2^0 \ra
\tilde{\tau}\tau$ increases.  The second class of processes listed
above is sensitive to violations of $e\mu$ universality only and
consequently $r_{SC}$ defined for same charge lepton pairs is
independent of violations of $e, \tau$ universality.

\begin{table}[t]
\caption{Ratios of production rates for lepton pairs for the five models studied (see text) 
\label{ratio}} 
\begin{center}
\begin{tabular}{cccccc} \hline \hline 
 $m_{3^{rd}}$ & 30 GeV & 70 GeV & 100 GeV (point 5) & 
150 GeV & 200 GeV \\  
\hline 
%$\frac{ee+\mu\mu}{e\mu}$ 
%& 2.28 & 3.26 & 3.54 & 3.92 &  3.73 \\  
$r_{OC} = \frac{e^+e^- + \mu^+\mu^-}{e^\pm\mu^\mp}$ 
& 2.61  & 3.86  & 3.99 &  4.62 &  4.38 \\
 & & & & & \\
$r_{SC} = \frac{e^\pm e^\pm + \mu^\pm\mu^\pm}{e^\pm\mu^\pm}$ 
& 0.88 & 0.79 & 1.25 & 1.05 & 1.01 \\
\hline \hline
\end{tabular}
\end{center}
\end{table}

\begin{figure}[t]
\dofig{3.5in}{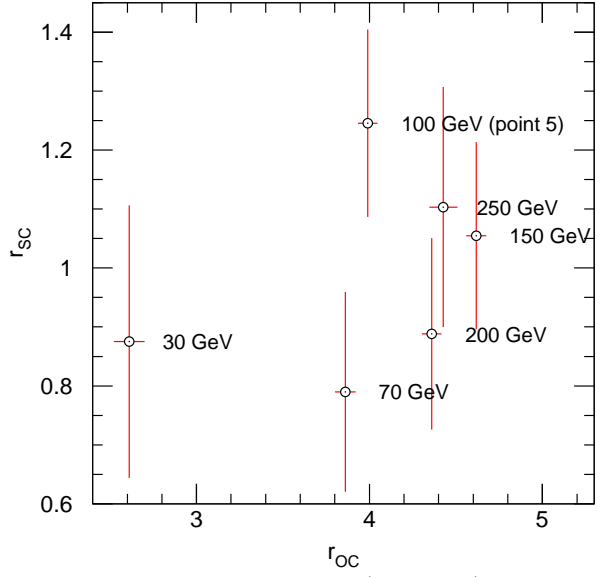}
\vskip-14pt
\caption{$r_{OC}$ and  $r_{SC}$ for various $m_{3^{rd}}$ 
values (see text). The error bars correspond to an integrated
luminosity of 10 fb$^{-1}$. 
\label{ratiogr}}
\end{figure}

Table~\ref{ratio} and Figure~\ref{ratiogr} show $r_{OC}$ and $r_{SC}$
for different values of $m_{3^{rd}}$ The error bars shown on
Figure~\ref{ratiogr} correspond to 10 fb$^{-1}$ of integrated
luminosity. $r_{SC}$ is insensitive to violations of $e/\mu$
universality and should be the same for all the cases considered. The
errors indicated on Figure~\ref{ratiogr} show that the apparent
differences are statistical fluctuations.

In the region between the benchmark case and $m_{3^{rd}} = 70\,\GeV$,
the contribution to $r_{OC}$ from the second class of processes
decreases dramatically since, at $m_{3^{rd}} = 70\,\GeV$, $\widetilde
\nu_\tau$ becomes significantly lighter than $\widetilde \chi_1^\pm$,
and the decay mode $\widetilde \chi_1^\pm \ra \tilde \nu_\tau \tau$
becomes important, with a branching ratio of 30\% instead of 0.4\%, to
the detriment of decays to electrons and muons.  For $m_{3^{rd}} \sim
75\,\GeV$, $\tilde \tau_2$ becomes lighter than $\tilde \chi_2^0$ (see
table~\ref{tau-mass}), and the decay channel $\tilde \chi_2^0 \ra
\tilde \tau_2 \tau$, opens resulting in a large violation of $e/\tau$
universality and increasing the number of OFOC pairs.  For $75\,\GeV <
m_{3^{rd}} < 100\,\GeV$, very slight change in the ratios is to be
expected, since these channels are not open, and the branching ratios
of the usual channel $\tilde \chi_2^0 \ra \tilde \ell \ell$ changes
only slowly.  For $m_{3^{rd}}>100$ GeV the only interesting decay mode
is $\tilde \chi_2^0 \ra \tilde \ell \ell$: as $m_{\tilde\tau_1}$
increases and tends to $m_{\tilde \chi_2^0}$, the decay branching ratio
$\tilde \chi_2^0 \ra \tilde \tau \tau $ closes, and $r_{OC}$ increases.
The channel closes at $m_{3^{rd}} = 200\,\GeV$.

To summarize: for $m_{3^{rd}} < 100\,\GeV$, $r_{OC}$ is very sensitive
 to $m_{3^{rd}}$. If $10\,\GeV < m_{3^{rd}} < 75\,\GeV$, one should be
able to constrain it with an accuracy of a few GeV.  For $m_{3^{rd}} >
100\,\GeV$, things are not so easy: for ($m_{3^{rd}} < 200\,\GeV$),
$r_{OC}$ increases slightly, giving some sensitivity to its value.  As
a result, models with $m_{3^{rd}}$ above 200 GeV can hardly be
distinguished from each other.

%%%%%%%%%%%%%%%%%%%%%%%%%%%%%%%%%%%%%%%%%%%%%%%%%%%%%%%%%%%%%%%%%%%%%%
\section{Determination of parameters}
\label{fits}

In previous studies\cite{HPSSY,P5} we needed to rely on a global fit to
a specific model, e.g., minimal SUGRA, to determine individual masses
from the combinations of masses measured by various end points. The
analysis in Section~\ref{ind-mass} allows us to extract masses in a
rather general way, but the global fit is still useful. Since the
measurements in Sections~\ref{four-body} and \ref{back-edge} provide new
information, we reevaluate here the precision with which the model
parameters can be determined. The strategy is the same as before: the
parameter space of the SUGRA model is searched using random sampling in
order to determine the $\pm34\%$ confidence limits resulting from the
assumed ``experimental'' quantities and their estimated errors.

The following quantities and estimated errors are used in the fit:
\arraycolsep=2pt % Stupid LaTeX
\begin{eqnarray*}
(M_{hq}^{\rm max})^2 &=& M_h^2
+ \left(M_{\tq}^2 - M_{\tchi_2^0}^2\right)
\left[{M_{\tchi_2^0}^2 + M_h^2 - M_{\lsp}^2 +
\sqrt{(M_{\tchi_2^0}^2 - M_h^2 - M_{\lsp}^2)^2 -4M_h^2
M_{\lsp}^2}} \over 2 M_{\tchi_2^0}^2 \right]\,,\\
&=& (552.6\pm40\,\GeV)^2\,;\\
M_{\ell\ell}^{\rm max} &=& M_{\tilde\chi_2^0}
\sqrt{1-{M_{\tilde\ell}^2 \over M_{\tilde\chi_2^0}^2}}
\sqrt{1-{M_{\lsp}^2 \over M_{\tilde\ell}^2}} = 108.9\pm0.1\,\GeV\,;\\
R = {M_{\ell q}^\rmax \over M_{\ell\ell q}^\rmax} &=&
\sqrt{\frac{M_{\tchi_2^0}^2-M_{\ell_R}^2}{M_{\tchi_2^0}^2-M_{\lsp}^2}}
=0.865\pm0.02\,;\\
M_{\ell q}^\rmax &=& \sqrt{ \frac{(M_{q_l}^2-M_{\tchi_2^0}^2)
(M_{\tchi_2^0}^2-M_{\ell_R}^2)}
{M_{\tchi_2^0}^2}} = 478.1\pm40\,\GeV \,;\\
(M_{\ell\ell q}^{\rm min})^2 &=& {1\over 4 M_2^2 M_e^2} 
\Biggl[-M_1^2 M_2^4  + 3 M_1^2 M_2^2 M_e^2 - M_2^4 M_e^2 - M_2^2 M_e^4 
- M_1^2 M_2^2 M_q^2 - \\
&&\quad M_1^2 M_e^2 M_q^2 + 3 M_2^2 M_e^2 M_q^2 - M_e^4 M_q^2 +
(M_2^2-M_q^2)\times \\
&&\quad \sqrt{(M_1^4+M_e^4)(M_2^2 + M_e^2)^2 +
2 M_1^2 M_e^2 (M_2^4 - 6 M_2^2 M_e^2 + M_e^4)}\Biggr]\,,\\
&=& (271.8\pm5.4\,\GeV)^2\,;\\
(M_{hq}^{\rm min})^2 &=& {1 \over 2M_2^2} (M_q^2-M_2^2) \times \\
&& \Bigl[(M_2^2+M_h^2-M_1^2)
-\sqrt{(M_2^2-M_h^2-M_1^2)^2-4M_1^2M_h^2}\Bigr]\,,\\
&=& (346.5\pm17\,\GeV)^2\,.
\end{eqnarray*} 
In addition we include $M_h$ in the fit with an error of $\pm3\,\GeV$.
The experimental error on the mass from $h \to \gamma\gamma$ will be
considerably less than this. The error reflects our estimate of the
theoretical uncertainty in relating the Higgs mass to the parameters of
the SUGRA model.

A fit of the minimal SUGRA model to these inputs results in the
following values of the parameters:
\begin{itemize}
\item $m_0=100.0 \pm 3.63$ GeV,
\item $m_{1/2}=300.0\pm 4.99$ GeV,
\item $\tan\beta=2.11 \pm 0.18$,
\item $\mu=+1$;
\end{itemize}
The errors are symmetric, unlike the earlier fits.  Recall that we
previously\cite{HPSSY} quoted:-
\begin{itemize}
\item $m_0=100^{+12}_{-8}$ GeV,
\item $m_{1/2}=300^{+6}_{-4}$ GeV,
\item $\tan\beta=1.8^{+0.3}_{-0.5}$,
\item $\mu=+1$;
\end{itemize}
Thus the new measurements improve the fit to the minimal SUGRA model as
well as allowing masses to be extracted without assuming the model.

A completely general model at the GUT scale would have as many
parameters as the MSSM. To keep the problem tractable, we consider three
variants of the SUGRA model, each with only one additional parameter. We
use the same fitting procedure to estimate how well these additional
parameters could be constrained if the actual data corresponded to the
benchmark case, { \it i.e.} we estimate how well we can actually
constrain to SUGRA model.

We first allow the values of $m_0$ at GUT scale to be different for the
two Higgs representations.  Restricting $m^2_{H_d} = m^2_{H_d} =
m_{h-GUT}>0$, leads to a 5-parameter fit and the following result ($m_0$
is now the common mass for all the other scalars)
\begin{itemize}
\item $m_0=100 \pm 3.68$ GeV,
\item $m_{1/2}=301\pm 5.94$ GeV,
\item $\tan\beta=2.11\pm .18$,
\item $\mu=+1$;
\item  $m_{h-GUT}< 430$ GeV, 95\% confidence
\end{itemize}
The insensitivity to $m_{h-GUT}$ arises because the derived value of
$\mu$ is large ($\sim 500$ GeV) and the value of $m_{H_U}$ (the Higgs
doublet that couples to charge 2/3 quarks) at the weak
scale is determined mainly by the top quark Yukawa coupling and stop
mass and not by the value of $m_{H_U}$ at the GUT scale ($m_{h-GUT}$).
  Therefore the
masses of $\lsp$, $\tchi_2^0$ and $h$ are insensitive to $m_{h-GUT}$ 
 unless it is very large. If the $H$ and $A$ Higgs bosons
(and the heaviest gauginos) could be observed and their masses measured
$m_{H_U}$ could be constrained.  The masses of these particles vary by
$\sim 40$ GeV for parameters in the allowed range.  The production rates
for these particles at LHC are extremely small and their discovery is
probably not possible there.  This insensitivity to $m_{h-GUT}$ is quite
general\cite{nick}.

We next split the squark and slepton masses at the unification scale:
the particles that are in the {\bf 10} of SU(5) are assumed to have
common scalar mass $m_{10}$ and those that lie in {\bf 5} of SU(5) are
assumed to have common scalar mass $m_5$.  Fitting for these, we get
\begin{itemize}
\item $m_{10}=100\pm 3.8$ GeV,
\item $m_{1/2}=300^{+10}_{-7}$ GeV,
\item $\tan\beta=2.11 \pm .23$,
\item $\mu=+1$;
\item  $m_5<420$ GeV, 95\% confidence
\end{itemize}

The lack of precise constraints on these new parameters can be
understood. Since $m_{1/2}$ is significantly larger than $m_0$, the
values of the squark masses at low energy are controlled by $m_{1/2}$.
The excellent constraint on $m_{10}$ arises because it controls
$m_{\ell_R}$ which is very precisely determined by the $\ell^+\ell^-$
end point. $\tilde{\ell_L}$ would be observable if the decay $\tchi_2^0
\to \tilde{\ell_L} \ell \to \ell^+\ell^- \lsp$\ were open as discussed
in Section~\ref{m5-down}. The failure to observe this decay then
constrains $m_{\ell_L}$ and consequently $m_5$.  Adding this constraint
excludes the region $m_5< 75 GeV$.  Direct production of right handed
sleptons excludes the region $m_5> 115$ GeV although this difficult
search requires high luminosity as explained in
Section~\ref{lept}. Using these constraints we infer:\\
\begin{itemize}
\item  $m_5=100^{+6}_{-10}$ GeV.
\end{itemize}

Finally, we consider the case where the third generation squark and
slepton masses at the GUT scale are allowed to vary. As can be seen
from Table~\ref{tau-mass}, the sensitivity is confined to the stau
masses so that a fit without the information from Section~\ref{thisrd}
results in almost no constraint on $m_{3rd}$. The errors on the other
parameters are as above. Adding the constraints
from this section implies that 
\begin{itemize}
\item $m_{3rd}=100^{-8}_{+4}$  GeV.
\end{itemize}

\section{Conclusions}
\label{conclude}

In this paper we have demonstrated a number of new techniques that
could be used to determine masses and decay properties of
supersymmetric particles at the LHC.  We have shown that cascade
decays with several steps can be used to reconstruct the masses of
supersymmetric partilces without any knowledge of the underlying model.
We have illustrated new signals that appear when the SUGRA model is
extended to have more parameters and have shown in particular how
$e/\tau$ universality could be tested. We have further demonstrated the
very high precision with which many of these parameters can be
constrained.

\section*{Acknowledgements}

This work was supported in part by the Director, Office of Science,
Office of Basic Energy Research, Division of High Energy Physics of
the U.S. Department of Energy under Contracts DE-AC03-76SF00098 and
DE-AC02-98CH10886.  Accordingly, the U.S.  Government retains a
nonexclusive, royalty-free license to publish or reproduce the
published form of this contribution, or allow others to do so, for
U.S. Government purposes. 

%%%%%%%%%%%%%%%%%%%%%%%%%%%%%%%%%%%%%%%%%%%%%%%%%%%%%%%%%%%%%%%%%%%%%%
\section*{Appendix: Attempt at complete reconstruction of SUSY events}

\addcontentsline{toc}{section}{Appendix: Attempt at complete
reconstruction of SUSY events}

\label{complete}

The three-step decay chain $\tq_L \to \tchi_2^0 \to \tell_R \to \lsp$
provides three mass constraints using the values determined in
Section~\ref{ind-mass} and so a $0C$ fit for the $\lsp$ momentum is
possible.  If the same decay is selected on both sides of the event,
then, in principle, one could completely reconstruct the event using
$\etmiss$ to select the best solution. This method was successful for
GMSB models\cite{HP} with decays involving leptons and photons, but it
fails in this case due to the experimental resolution, as we will now
show.

Events were selected to be consistent with two $\tchi_2^0 \to
\tell^\pm\ell^\mp \to \lsp\ell^+\ell^-$ decays and so to have four
leptons, at least two jets, and missing energy:
\begin{itemize}
\item   $\Meff > 400\,\GeV$.
\item   $\etmiss > \max(0.2\Meff,100\,\GeV)$.
\item   At least two  jets with $p_{T,1}>100\,\GeV$, $p_{T,2}>75\,\GeV$ and
at least two charged tracks in $R=0.4$.
\item   Four isolated leptons with $p_T>10\,\GeV$, $\eta<2.5$.
Isolation being defined so that there is  less than 
10 GeV of additional transverse energy in a cone  $R=0.2$ around the lepton direction.
\end{itemize}
The cut on the charged multiplicity of the jets was made to eliminate
electrons and hadronic tau decays from the jet sample.  It was further
required that there be one and only one way to form two opposite-sign,
same-flavor pairs of the four leptons with $20\,\GeV < M_{\ell\ell} <
115\,\GeV$. This ensures an unambiguous paring of the leptons consistent
with two $\tchi_2^0$ decays.

\begin{figure}[t]
\dofig{3.5in}{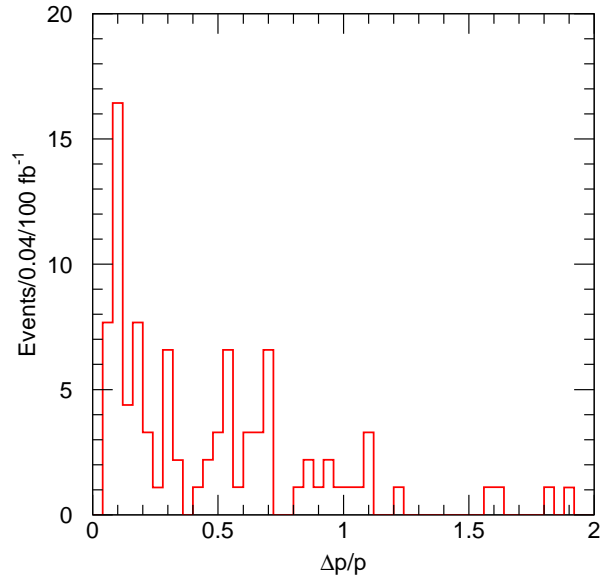}
\caption{difference $\Delta p/p$ between the generated
and the best reconstructed $|\vec p|$. \label{c5_dpp}}
\end{figure}

The events were then fit to the hypothesis that the four leptons and the
two highest $p_T$ jets came from the $\tq_L \ra \tchi_2^0 \ra \tell \ra
\lsp$ decay chain. For each such decay chain there are three mass
constraints, namely
\begin{eqnarray*}
(p_\lsp + p_{\ell,1})^2 &=& M_{\tell}^2 \\
(p_\lsp + p_{\ell,1} + p_{\ell,2})^2 &=& M_{\tchi_2^0}^2 \\
(p_\lsp + p_{\ell,1} + p_{\ell,2} + p_q)^2 &=& M_{\tq}^2
\end{eqnarray*}
There are also two constraints from $\etmiss$. Since the measurement
errors on the jets are comparable to those on $\etmiss$, the jet
energies were smeared by factors $\lambda_i$ distributed in Gaussian
manner, and the best solution was taken to be the one which minimizes
$$
\chi^2 
= {\left(\slashchar{E}_x - p_{1x} - p_{2x}\right)^2 \over 
\sigma^2(\slashchar{E}_x)}
+ {\left(\slashchar{E}_y - p_{1y} - p_{2y}\right)^2 \over 
\sigma^2(\slashchar{E}_y)}
+ {\lambda_1^2 \over \sigma^2(\lambda_1)}
+ {\lambda_2^2 \over \sigma^2(\lambda_2)}\,,
$$
where both the missing energy resolutions
$\sigma(\slashchar{E}_{x,y})$ and the jet energy scale resolutions
$\sigma(\lambda_{1.2})$ are determined using the Gaussian calorimeter
resolution. The resulting difference $\Delta p/p$ between the
generated and the best reconstructed $|\vec p|$ for the $\lsp$ is
shown in Figure~\ref{c5_dpp}.  A similar reconstruction was successful
in the case of the GMSB study\cite{HP} where the decay chain 
$$
\tchi_2^0 \to \tell^\pm\ell^\mp \to \lsp\ell^\pm\ell^\mp \to
\tG\gamma\ell^\pm\ell^\mp
$$
was used.  The reconstruction works much more poorly than in the GMSB
case; while a few events are correctly reconstructed, most are not.
This is not very surprising; the GMSB case relied on leptons and photons
which have much better energy resolution than the jets used here. In
addition, in this case, the $\lsp$ momenta are significantly smaller
than the jet momenta and so the errors on the jet energy measurements
are very important.

\end{document}